\documentclass[12pt]{article}

\usepackage{amssymb, amsmath, amsthm, mathtools, commath}
\usepackage{mathrsfs, bbm, bm}

\usepackage{graphicx, subcaption, xcolor}

\usepackage{booktabs, array, float, multirow, tabularx, longtable, arydshln}

\usepackage[T1]{fontenc}
\usepackage[utf8]{inputenc}
\usepackage{fullpage}
\usepackage{makeidx}
\usepackage{comment}
\usepackage{enumitem}
\usepackage[round]{natbib}
\usepackage[unicode,hidelinks]{hyperref}

\DeclareMathOperator*{\argmin}{arg\,min}
\DeclareMathOperator{\Var}{Var}
\DeclareMathOperator{\PP}{P}
\DeclareMathOperator{\MSE}{MSE}

\theoremstyle{plain}

\theoremstyle{definition}

\newcommand{\T}{\top}
\newcommand{\bx}{\bm{x}}

\newcommand{\bz}{\bm{z}}

\newcommand{\balpha}{\bm{\alpha}}
\newcommand{\bbeta}{\bm{\beta}}

\newcommand{\btheta}{\bm{\theta}}

\newcommand{\bgamma}{\bm{\gamma}}
\newcommand{\br}{\bm{r}}

\newcommand{\bC}{\bm{C}}

\newenvironment{keywords}{%
  \par\noindent\textbf{Key words:}\ }{%
  \par\medskip}

\numberwithin{equation}{section}
\numberwithin{table}{section}
\numberwithin{figure}{section}
\newcommand{\tabnote}[1]{%
  \par\vspace{2mm}\begin{minipage}{0.95\linewidth}\footnotesize\raggedright\textit{Note:} #1\end{minipage}%
}
\newcommand{\fignote}[1]{%
  \par\vspace{2mm}\begin{minipage}{0.95\linewidth}\footnotesize\raggedright\textit{Note:} #1\end{minipage}%
}
\hypersetup{
  pdftitle={Small area estimation using incomplete auxiliary information},
  pdfauthor={Donatas Šlevinskas, Ieva Burakauskaitė, Andrius Čiginas}
}

\begin{document}

\title{\textbf{Small area estimation using incomplete auxiliary information}}

\author{
Donatas~Šlevinskas \qquad
Ieva~Burakauskaitė \qquad
Andrius~Čiginas\thanks{Corresponding author. E-mail: \texttt{andrius.ciginas@mif.vu.lt}.}\\[0.4em]
{\small Vilnius University, Faculty of Mathematics and Informatics,}\\
{\small Institute of Data Science and Digital Technologies}\\
{\small Akademijos str.~4, LT-08412 Vilnius, Lithuania}
}

\date{}

\maketitle

\begin{abstract}
Auxiliary information is increasingly available from administrative and other data sources, but it is often incomplete and of non-probability origin. We propose a two-step small area estimation approach in which the first step relies on design-based model calibration and exploits a large non-probability sample providing a noisy proxy of the study variable for only part of the population. A unit-level measurement-error working model is fitted on the linked overlap between the probability survey and the non-probability sample, and its predictions are incorporated through domain-specific model-calibration constraints to obtain approximately design-unbiased domain totals. These totals and their variance estimates are then used in a Fay--Herriot area-level model with exactly known covariates to produce empirical best linear unbiased predictors. The approach is demonstrated in three enterprise survey settings from official statistics by integrating probability sample data with (i) administrative records, (ii) a cut-off sample, and (iii) web-scraped online information. Empirical comparisons show consistent improvements in domain-level precision over direct estimation and over a Fay--Herriot benchmark that directly incorporates the proxy information as an error-prone covariate. These gains are achieved without modeling the mechanism that generates the non-probability sample: the realized sample is used as fixed auxiliary information, while the working prediction function is estimated from the probability sample under the sampling design.
\end{abstract}


\section{Introduction}\label{sec:introduction}

Probability sample surveys are the cornerstone for producing reliable estimates of population parameters in survey statistics, including official statistics. Yet, survey data alone are often insufficient for small population domains, where direct design-based estimators can have large variances due to small domain sample sizes. At the same time, alternative data sources, such as administrative registers or big data, are increasingly available.
These sources are rarely complete and are often of non-probability origin, but they may provide valuable auxiliary information closely related to the study variables.

We develop a design-based method that uses incomplete auxiliary information to improve domain-level estimation. The resulting domain estimates can subsequently be modeled within small area estimation (SAE) frameworks. Our work is motivated by three applications from official statistics on enterprises in Lithuania, where we estimate the totals of
\begin{enumerate}[topsep=3pt, itemsep=2pt]
    \item \textit{monthly turnover} of enterprises, where closely related value-added tax (VAT) data are available only for part of the population;
    
    \item \textit{annual investment in tangible assets}, where a cut-off sample from an alternative structural business statistics survey provides partial information closely aligned with the study concept;
    
    \item \textit{quarterly job vacancies}, where online job advertisements (OJA) data are obtained from non-probability web scraping.
\end{enumerate}
These cases share a common challenge: additional information exists and is closely related to the study variable, but it is incomplete and subject to unknown selection mechanisms, and may differ from the study variable due to alternative definitions or measurement errors. This setting makes classical calibration or model-assisted estimation methods inapplicable in their standard form.

From the perspective of non-probability sampling, our setting is best understood in terms of the four-type classification summarized by \cite{TH_2020}. Table~\ref{tab:types} summarizes the main types. Our work aligns with Type IV, where proxy measurements of the study variable are available only for a subset of the population through a non-probability sample and can be incorporated as auxiliary information. This case has received limited attention, with \cite{MGRBDP_2024} covering only the linear model case. By contrast, our approach allows for substantially more general relationships between the observed proxies and the study variables. The setup is also closely related to the Type III data integration of \cite{KT_2021}, which applies design-based estimation methods; however, our formulation extends beyond their linear measurement-error model. For comparison, Type II approaches -- mass imputation, propensity-based inverse probability weighting (IPW), and doubly-robust adjustments using a supplementary probability sample \citep{EV_2017, CLW_2020, KPCW_2021} -- are widely considered in the literature but typically rely on an unverifiable missing-at-random (MAR) assumption for the non-probability sample selection mechanism. In our Type IV setting, the proposed estimator uses the realized non-probability sample as fixed auxiliary information for the subset of the finite population in which the proxy is observed rather than as a representative sample for the whole population. The probability sample remains the basis for design-based inference, and the linked overlap is used to estimate the working prediction function under the sampling design. As a result, the approach avoids specifying a MAR-type model for the selection mechanism of the non-probability sample; its design-based properties rely on the model-calibration conditions discussed in Section~\ref{sec:methodology}.

\begin{table}[!ht]
    \centering
    \caption{Types of data source structure.}
    \begingroup
    \small
    \setlength{\tabcolsep}{4pt}
    \begin{tabular}{llccc}
        \toprule 
        & \textbf{Source} & \begin{tabular}{c} 
        \textbf{Study} \\
        \textbf{variable $y$}
        \end{tabular} & \begin{tabular}{c} 
        \textbf{Auxiliary} \\
        \textbf{variables $\bx$}
        \end{tabular} & \begin{tabular}{c} 
        \textbf{Design} \\
        \textbf{weights}
        \end{tabular} \\
        \toprule Type I & Non-probability sample, $B$ & $\checkmark$ & $\checkmark$ \\
        \midrule \multirow{2}{*}{Type II} 
            & Non-probability sample, $B$ & $\checkmark$ & $\checkmark$ & \\
            & Probability sample, $A$ &  & $\checkmark$ & $\checkmark$ \\
        \midrule \multirow{2}{*}{Type III} 
            & Non-probability sample, $B$ & $\checkmark$ & $\checkmark$ \\
            & Probability sample, $A$ & $\checkmark$ & $\checkmark$ & $\checkmark$\\
        \midrule \multirow{2}{*}{Type IV} 
            & Non-probability sample, $B$ & & $\checkmark$ \\
            & Probability sample, $A$ & $\checkmark$ & $\checkmark$ & $\checkmark$ \\ 
        \bottomrule
    \end{tabular}
    \endgroup
    \label{tab:types}
\end{table}

We develop a methodology that combines probability sample data with incomplete auxiliary information, following the model-calibration framework of \citet{WS_2001} and adapting ideas from \citet{KT_2021} on data integration. The proposed estimators are approximately design-unbiased and can be further improved by applying SAE models.

SAE methodology has been developed in two main directions: unit-level models (e.g., \cite{BHF_1988}) and area-level models (e.g., \cite{FH_1979}); a comprehensive account is given by \cite{RM_2015}. In our framework, the model-calibration step can be viewed as producing domain-level estimators that already incorporate unit-level auxiliary information observed only for a subset of units. Consequently, subsequent modeling naturally proceeds at the area level, where, for example, the Fay--Herriot (FH) model borrows strength across domains using covariates based on completely known auxiliary data (i.e., exactly known or error-free area-level covariates). 

The contributions of this paper are twofold. First, we extend model calibration to the Type IV setting by estimating, under the probability sampling design, a general working relationship between the study variable and a proxy observed only in the non-probability sample, and by incorporating the resulting fitted auxiliary total through domain-specific calibration constraints. The resulting calibrated estimators can then serve as inputs to SAE models, further reducing estimation uncertainty. Second, we demonstrate the approach through applications to Lithuanian official statistics and compare the resulting estimates with direct estimators and with an area-level benchmark adapted from \citet{YL_2008}.

The remainder of the paper is organized as follows.
Section~\ref{sec:methodology} introduces the general setup, develops the model-calibration framework, and describes how the resulting estimators are incorporated into the SAE stage. Here, we also present an adaptation of the method by \cite{YL_2008}. Section~\ref{sec:results} presents results from three empirical applications in Lithuanian official statistics: monthly turnover, annual investment in tangible assets, and quarterly job vacancies. Section~\ref{sec:simulation} examines design bias, working-model misspecification, departures from MAR, and the role of domain-level overlap in a focused design-based simulation.
Section~\ref{sec:conclusions} concludes with methodological and practical implications for integrating probability and non-probability samples.

\section{Methodology}\label{sec:methodology}

\subsection{General setup}

We begin by defining the finite population, sampling design, and notation used throughout the paper. Let $U=\{1,\ldots,N\}$ denote a finite population of size $N$, partitioned into $M$ non-overlapping domains $U=U_1\cup\cdots\cup U_M$, where $U_m$ has size $N_m$. A probability sample $A\subset U$ of size $n$ is drawn according to a known sampling design with first-order inclusion probabilities $\pi_i>0$ for $i\in U$, with corresponding design weights $d_i=1/\pi_i$ used to form design-based estimators. The values $y_i$ of the study variable $y$ are observed for all $i\in A$. For each domain $m=1,\ldots,M$, let $A_m=A\cap U_m$ denote the part of the probability sample belonging to $U_m$.

The aim is to estimate the domain totals
\begin{equation*}
t_m=\sum_{i\in U_m} y_i, \quad m=1,\ldots,M.
\end{equation*}
If the domain-specific sample sizes $n_m=|A_m|$ are small, direct design-based estimators such as the Hájek estimators
\begin{equation}\label{eq:dir_H}
\hat{t}_m^{\mathrm H}=\frac{N_m}{\hat{N}_m}\sum_{i\in A_m} d_i y_i, \quad \text{where }\hat{N}_m=\sum_{i\in A_m} d_i,
\end{equation}
can yield unreliable estimates for small areas. 

In many official statistics applications, auxiliary variables $\bx_i=(1, x_{i1},\ldots,x_{ip})^\T$, $p\geq 1$, are available for nearly all population units from administrative registers, censuses, or other external sources. Such information can be used to improve the direct estimators, such as \eqref{eq:dir_H}, through the standard calibration approach of \cite{DS_1992} or generalized regression (GREG) estimation, leading to approximately design-unbiased estimators with smaller variances at the domain level.

In addition to these fully available auxiliary variables, other data sources may provide further information more closely related to the study variable. Suppose that a large non-probability sample $B\subset U$ of size $N_B$ is available and overlaps with the probability sample $A$. In the design-based argument below, the realized sample $B$ is treated as fixed auxiliary information in the finite population. We assume that the records of $A$ and $B$ can be linked through common identifiers. For units $i\in B$, we observe the values $y_i^\star$ of a proxy variable $y^\star$ that represents a noisy version of the study variable $y$ -- for example, measurements based on administrative or alternative sources with definitions that differ from the study concept.

The intersection $A\cap B$ contains units for which both variables $y$ and $y^\star$ are observed. Within the fixed $B$-post-stratum, the relationship between $y_i$ and $y_i^\star$ is represented by a general measurement-error working model of the form
\begin{equation}\label{eq:me_model}
y_i=g(y_i^\star,\bx_i)+\varepsilon_i,\quad i\in B,
\end{equation}
where $g(\cdot)$ is a working prediction function, linear, nonlinear, or nonparametric depending on the data structure, and $\varepsilon_i$ is a working residual. The overlap $A\cap B$ provides the observations used to estimate this working relationship under the probability sampling design. For finite-dimensional models, the target relationship can be expressed as $E(\varepsilon_i\mid y_i^\star,\bx_i,i\in B)=0$ with finite conditional variance, but no full distributional model for $\varepsilon_i$ is required for the model-calibration argument. The empirical examples mentioned in Section~\ref{sec:introduction} and further studied in Section~\ref{sec:results} mainly differ by the assumed form of $g(\cdot)$. The fitted working model provides values $\hat y_i$ for units in $B$. These fitted values act as an additional auxiliary variable in the $B$-part of each domain, while the part outside $B$ enters the calibration through a separate post-stratification constraint introduced below. This framework enables the use of incomplete auxiliary information to construct approximately design-unbiased estimators of domain totals through the model-calibration approach described in the next subsection.

\subsection{Model-calibration framework}

Following \cite{KT_2021}, we divide the population into two artificial strata (post-strata): the $B$-post-stratum, where the proxy $y^\star$ is observed, and the missing-data post-stratum, where it is not. The set $B$ is thus used as an observed post-stratification partition of the finite population. Let the variable
\begin{equation*}
\delta_i =
\begin{cases}
1, & \text{if } i \in B,\\
0, & \text{otherwise},
\end{cases}
\end{equation*}
indicate the availability of $y_i^\star$. The probability sample $A$ contains units from both strata, and we assume that, for each domain $U_m$, the corresponding sample part $A_m$ includes sufficient information from both post-strata. In practice, this means that both post-strata are represented in $A_m$, that the $B$-part provides enough variation in the fitted values $\hat y_i$ to support the count and fitted-total constraints, and that the resulting calibration weights are not extreme.

Let $\hat{y}_i=\hat{g}(y_i^\star,\bx_i)$ denote the predicted value of $y_i$ obtained from the measurement-error model \eqref{eq:me_model} fitted using units in $A\cap B$. For finite-dimensional working models $g(\cdot;\btheta)$, let $\btheta_B$ denote the finite-population parameter in the fixed $B$-post-stratum, defined as the solution of
\begin{equation*}
\sum_{i\in B}\psi(y_i,y_i^\star,\bx_i;\btheta_B)=0,
\end{equation*}
where $\psi(\cdot)$ is the estimating function corresponding to the chosen working model. Since $y_i$ is observed only for units in the probability sample, $\btheta_B$ is estimated from the overlap by the design-weighted estimating equation
\begin{equation*}
\sum_{i\in A\cap B} d_i\,\psi(y_i,y_i^\star,\bx_i;\hat\btheta_B)=0.
\end{equation*}
The fitted auxiliary values are then $\hat y_i=g(y_i^\star,\bx_i;\hat\btheta_B)$, $i\in B$. For nonparametric smoothers, such as the $k$-nearest neighbors ($k$NN) predictor used below, the same design-based principle is implemented by using the design weights as case weights in the local averaging step. In the model-assisted sense of \cite{WS_2001}, this prediction function serves as a working model whose correctness is not required for subsequent design-based inference.

For each domain $U_m$, let $d_i$ denote the baseline weights used in calibration; these may be either the original design weights or other design-consistent weights, such as those obtained from a prior GREG adjustment. The calibration weights $w_i$, $i\in A_m$, are required to satisfy the following domain-specific constraints:
\begin{align}
\sum_{i\in A_m} w_i (1-\delta_i) &= N_m - N_{B_m}, \label{eq:mc_constr1}\\
\sum_{i\in A_m} w_i \delta_i &= N_{B_m}, \label{eq:mc_constr2}\\
\sum_{i\in A_m} w_i \delta_i \hat{y}_i &= \sum_{i\in B_m} \hat{y}_i, \label{eq:mc_constr3}
\end{align}
where $N_{B_m}=|B_m|$ and $B_m=B\cap U_m$. The first two constraints are post-stratification constraints matching the population sizes of the missing-data and $B$-post-strata within each domain, while the third is the model-calibration constraint aligning the fitted auxiliary total over $B_m$.

The calibration weights $w_i$ are obtained by minimizing the chi-square distance \citep{DS_1992}
\begin{equation*}
\sum_{i\in A_m} d_i \left(\frac{w_i}{d_i}-1\right)^2,
\end{equation*}
subject to the constraints \eqref{eq:mc_constr1}--\eqref{eq:mc_constr3}, following the model-calibration framework of \cite{WS_2001}. The resulting model-calibration estimator of the domain total is
\begin{equation}\label{eq:mc_domain}
\hat{t}_m^{\mathrm{MC}} = \sum_{i\in A_m} w_i y_i, \quad m=1,\ldots,M.
\end{equation}

Under regularity conditions of the type considered by \cite{WS_2001}, including design-consistent estimation of the working model parameters, nonsingular calibration matrices, and bounded moments, the estimator \eqref{eq:mc_domain} is approximately design-unbiased for $t_m$. The correctness of $g(\cdot)$ is not required for this design-based property; misspecification mainly affects the efficiency gain obtained from calibration. The estimator is expected to be more efficient than the direct estimator using the baseline weights when the proxy values $y_i^\star$ (and thus the predictions $\hat{y}_i$) are strongly related to $y_i$. If either post-stratum is absent from $A_m$, the calibration equations are rank-deficient, or the implied weights become unstable, the calibration system \eqref{eq:mc_constr1}--\eqref{eq:mc_constr3} is not used; the direct estimator and its design-based variance are used instead as first-stage inputs to the FH model, as detailed in Section~\ref{sec:FH}. 

Building on \cite{DS_1992} and \cite{WS_2001} and extending the data-integration framework of \cite{KT_2021}, this formulation allows for incomplete auxiliary information and general, possibly nonlinear, relationships between the proxy and the study variable, enabling design-consistent integration of probability and non-probability samples at the domain level.

\subsection{Variance estimation}

The variance of the model-calibration estimator $\hat{t}_m^{\mathrm{MC}}$ given by \eqref{eq:mc_domain} can, in principle, be approximated by linearization as in \citet{WS_2001}. However, such asymptotic approximations are often unreliable for small domains, where sample sizes $n_m$ are limited and the estimation procedure involves nonlinear adjustments. A more robust and broadly applicable alternative is to estimate the variance by resampling, using the design-based bootstrap of \citet{RWY_1992}. This approach produces variance estimators that are consistent under the sampling design and does not require any assumptions about the correctness of the working model behind the calibration adjustment.

In our applications considered in Section~\ref{sec:results}, the survey data are obtained from stratified simple random samples without replacement, with strata defined by enterprise characteristics such as economic activity and size class. The Rao--Wu--Yue bootstrap was developed for general stratified multistage designs and can be directly applied to stratified simple random sampling. Since the fitted working prediction function $\hat g$ is estimated from $A\cap B$, it is random under the sampling design. We therefore use a bootstrap procedure that re-estimates the working prediction function in each replicate using the Rao--Wu--Yue bootstrap weights. This incorporates the design variability due to both the estimation of $\hat g$ and the subsequent calibration step, while retaining the model-assisted interpretation of $g(\cdot)$ as a working model.

The bootstrap variance estimator is obtained by repeating the complete estimation procedure within each resampled dataset. The steps are as follows:

\begin{enumerate}[label=(\alph*)]
\item Let the population be divided into $H$ strata, and let $A_h$ denote the part of the probability sample $A$ selected from stratum~$h$ by simple random sampling without replacement of size $n_h$. For each stratum $h=1,\ldots,H$, draw a bootstrap sample $A_h^{*(r)}$ of size $n_h-1$ by simple random sampling with replacement from~$A_h$, independently across strata.

\item For each unit $i\in A_h$, record the number of times $m_{hi}^{*(r)}$ that unit $i$ is selected into $A_h^{*(r)}$, and compute the corresponding bootstrap design weight as
\begin{equation}\label{boot_w}
d_{hi}^{*(r)} = \frac{n_h}{n_h - 1} m_{hi}^{*(r)} d_{hi}.
\end{equation}
The replicate can equivalently be represented by the original sampled units with weights $d_{hi}^{*(r)}$, with zero weight assigned to units not selected in the bootstrap draw. Let $A^{*(r)}=\{i\in A:d_i^{*(r)}>0\}$ and $A_m^{*(r)}=A^{*(r)}\cap U_m$.

\item Re-estimate the working prediction function using the bootstrap design weights on the linked overlap $A\cap B$. For finite-dimensional working models, obtain $\hat\btheta_B^{*(r)}$ from
\begin{equation*}
\sum_{i\in A\cap B} d_i^{*(r)}\,\psi(y_i,y_i^\star,\bx_i;\hat\btheta_B^{*(r)})=0,
\end{equation*}
and set
\begin{equation*}
\hat y_i^{*(r)}=g(y_i^\star,\bx_i;\hat\btheta_B^{*(r)}),\quad i\in B.
\end{equation*}
For nonparametric predictors such as the $k$NN smoother, recompute the local predictions using $d_i^{*(r)}$ as case weights for the training units in $A\cap B$.

\item For each domain $U_m$, determine the bootstrap calibration weights $w_i^{*(r)}$ by minimizing the same chi-square distance as in the model-calibration estimator, using $d_i^{*(r)}$ as the baseline weights and imposing the replicate calibration constraints
\begin{align*}
\sum_{i\in A_m^{*(r)}} w_i^{*(r)} (1-\delta_i) &= N_m - N_{B_m}, \\
\sum_{i\in A_m^{*(r)}} w_i^{*(r)} \delta_i &= N_{B_m}, \\
\sum_{i\in A_m^{*(r)}} w_i^{*(r)} \delta_i \hat{y}_i^{*(r)} &= \sum_{i\in B_m} \hat{y}_i^{*(r)}.
\end{align*}

\item Obtain the bootstrap replicate of the domain estimator:
\begin{equation*}
\hat{t}_m^{\mathrm{MC}*(r)} = \sum_{i\in A_m^{*(r)}} w_i^{*(r)} y_i, \quad m=1,\ldots,M.
\end{equation*}

\item Repeat steps (a)--(e) for $r=1,\ldots,R$ bootstrap replicates and estimate the variance as
\begin{equation*}
\begin{aligned}
\hat{V}_{\mathrm{RWY}}(\hat{t}_m^{\mathrm{MC}})
&=\frac{1}{R-1}\sum_{r=1}^{R}\left(\hat{t}_m^{\mathrm{MC}*(r)} - \bar{t}_m^{\mathrm{MC}*}\right)^2,\\
\bar{t}_m^{\mathrm{MC}*}
&=\frac{1}{R}\sum_{r=1}^{R}\hat{t}_m^{\mathrm{MC}*(r)}.
\end{aligned}
\end{equation*}
\end{enumerate}

In bootstrap replicates where model calibration was not applied, a direct estimate was used as the fallback.

The non-probability sample $B$ is kept fixed in the bootstrap replicates: its records, domain membership and proxy values are not resampled, and the fitted auxiliary totals on the right-hand side of the calibration constraints are computed over the same fixed sets $B_m$. Only the bootstrap weights attached to units of the probability sample change; hence units in $A\cap B$ enter the re-estimation of $\hat g^{*(r)}$ through their replicate weights and original linked identifiers. In this context, reconstructing a pseudo-population as in \citet{BBH_1994} could distort the structure of~$B$, since its composition would then depend on the resampling weights of $A$. Therefore, the Rao--Wu--Yue bootstrap provides a coherent design-based resampling method in the integrated $(A,B)$ data setting, yielding stable and design-consistent variance estimates.

The resulting bootstrap variance estimator $\hat{V}_{\mathrm{RWY}}(\hat{t}_m^{\mathrm{MC}})$ can be used directly to assess the reliability of the model-calibration estimates. If the resulting variances are sufficiently small, no further modeling may be necessary. Otherwise, the estimates $\hat{t}_m^{\mathrm{MC}}$ and their bootstrap variances serve as inputs to the area-level modeling step described in the following subsection.

\subsection{Area-level modeling}\label{sec:FH}

In business and economic surveys, domain totals, such as those of turnover, investment, or job vacancies in our applications of Section~\ref{sec:results}, are typically positive and right-skewed, with variances increasing as the level of the aggregated study variable increases. In such cases, the assumptions of linearity and normality in the basic FH model are often violated. A log-transformation of the first-stage domain estimators helps stabilize variances and better satisfy these model assumptions.

Let $\mathcal M_{\mathrm{FH}}$ denote the set of domains included in the FH fit. For each $m\in\mathcal M_{\mathrm{FH}}$, define the first-stage estimator
\begin{equation*}
\hat t_m^{(1)}=
\begin{cases}
\hat t_m^{\mathrm{MC}}, & \text{if the model-calibration estimator is used in domain }m,\\
\hat t_m^{\mathrm d}, & \text{otherwise},
\end{cases}
\end{equation*}
where $\hat t_m^{\mathrm d}$ is the corresponding direct estimator. Domains outside $\mathcal M_{\mathrm{FH}}$, including exact take-all domains, receive no FH prediction; their direct estimators are retained unchanged as the final domain estimates. Let $\hat t_m^{(1,\log)}=\log(\hat t_m^{(1)})$ and denote its sampling variance on the log scale by $\psi_m^{(1,\log)}$. For domains where the model-calibration estimator is used, this variance is estimated using the Rao--Wu--Yue replicates described in Section~2.3, thereby propagating the variability from re-estimating the working prediction function and calibration. For the remaining domains in $\mathcal M_{\mathrm{FH}}$, the design-based variance estimator of the direct estimator on the original scale is converted to the log scale using the delta method. We denote the resulting variance estimator by $\hat{\psi}_m^{(1,\log)}$.

The FH model on the log scale is written as
\begin{equation}\label{eq:FH_log}
\begin{aligned}
\hat{t}_m^{(1,\log)} &= t_m^{(\log)} + e_m, 
& e_m &\overset{\text{ind}}{\sim}\mathcal{N}(0,\psi_m^{(1,\log)}),\\
t_m^{(\log)} &= \bz_m^{\T}\bbeta + u_m, 
& u_m &\overset{\text{ind}}{\sim}\mathcal{N}(0,\sigma_u^2),
\end{aligned}
\end{equation}
where $t_m^{(\log)}=\log(t_m)$ denotes the true but unobserved log-transformed domain total, $\bz_m=(z_{m0}, z_{m1}, \ldots, z_{mq})^\T$, $q\leq p$, are fixed domain-level covariates obtained as functions of completely known auxiliary data $\bx_i$, $i\in U$, the vector parameter $\bbeta$ represents fixed effects, and $u_m$ are area-specific random effects assumed independent of the sampling errors $e_m$.

Treating the variances $\hat{\psi}_m^{(1,\log)}$ as known when estimating the parameters $\bbeta$ and $\sigma_u^2$, the empirical best linear unbiased predictor (EBLUP) of $t_m^{(\log)}$ is
\begin{equation*}\label{eq:EBLUP_log}
\tilde{t}_m^{(\log)}=\hat{\gamma}_m\hat{t}_m^{(1,\log)}+(1-\hat{\gamma}_m)\bz_m^{\T}\hat{\bbeta} \quad \text{with } \hat{\gamma}_m=\frac{\hat{\sigma}_u^2}{\hat{\sigma}_u^2+\hat{\psi}_m^{(1,\log)}}
\end{equation*}
and
\begin{equation*}\label{eq:beta_hat}
\hat{\bbeta}=\left(\sum_{m\in\mathcal M_{\mathrm{FH}}} \frac{\bz_m\bz_m^{\T}}{\hat{\sigma}_u^2+\hat{\psi}_m^{(1,\log)}}\right)^{-1}\sum_{m\in\mathcal M_{\mathrm{FH}}} \frac{\bz_m\hat{t}_m^{(1,\log)}}{\hat{\sigma}_u^2+\hat{\psi}_m^{(1,\log)}}.
\end{equation*}
To obtain predictions on the original scale, the log-normal bias correction is applied:
\begin{equation}\label{eq:EBLUP_back}
\hat{t}_m^{\mathrm{EBLUP}}=\exp\left\{\tilde{t}_m^{(\log)}+0.5 \, \widehat{\mathrm{MSE}}(\tilde{t}_m^{(\log)})\right\}, \qquad m\in\mathcal M_{\mathrm{FH}},
\end{equation}
where $\widehat{\mathrm{MSE}}(\tilde{t}_m^{(\log)})$ denotes the estimated mean squared error (MSE) of the EBLUP on the log scale. The MSE can be evaluated by standard analytical approximations, depending on the method used to estimate the random effects variance $\sigma_u^2$ (such as restricted maximum likelihood (REML) or moment-based estimators), or, preferably, by a parametric bootstrap under model~\eqref{eq:FH_log}, which provides a flexible way to account for the estimation uncertainty in both $\hat{\sigma}_u^2$ and $\hat{\psi}_m^{(1,\log)}$, see, e.g., Ch.~6 in \cite{RM_2015}. The MSE for the back-transformed predictor \eqref{eq:EBLUP_back} can be evaluated accordingly, using either the delta method or a parametric bootstrap that reproduces both the estimation and back-transformation steps.

The adequacy of the log-normal approximation underlying \eqref{eq:EBLUP_back} is assessed in Section~\ref{sec:app_estim} using standardized marginal residuals from model~\eqref{eq:FH_log}.

The EBLUPs $\hat{t}_m^{\mathrm{EBLUP}}$ thus provide model-based improvements over the first-stage estimators $\hat t_m^{(1)}$, borrowing strength across domains through the auxiliary error-free covariates $\bz_m$.

\subsection{Fay--Herriot model with an error-prone auxiliary total}

As an alternative to the model-calibration estimation combined with the classical EBLUP, we consider a modification of the FH model that accounts for measurement error in one auxiliary area-level total. The motivating setting is that closely related additional information on the study variable is available from the non-probability sample $B$, but the corresponding population total is unknown. In this case, we estimate these totals using IPW estimators and then incorporate them into the FH model, following the approach of \citet{YL_2008}. In the empirical comparisons, this adaptation serves as an area-level benchmark that summarizes the proxy information through an estimated domain total.

\subsubsection{Inverse probability weighted estimator for auxiliary totals}

Let us consider the domain totals
\begin{equation*}
t_{m\star}=\sum_{i\in U_m} y_i^\star, \quad m=1,\ldots,M,
\end{equation*}
of the auxiliary variable $y^\star$, which is observed only for units in the non-probability sample~$B$. We estimate these totals using the IPW estimators
\begin{equation}\label{IPW_est}
\hat t^{\mathrm{IPW}}_{m\star}=\frac{N_m}{\hat N_{B_m}} \sum_{i\in B_m} \frac{y_i^{\star}}{\hat\pi_i} \quad \text{with } \hat N_{B_m}=\sum_{i\in B_m}\frac{1}{\hat\pi_i}.
\end{equation}
The estimated propensity scores $\hat\pi_i=\pi(\bx_i,\hat\btheta)$ are obtained by fitting the logistic regression model
\begin{equation*}
\PP(\delta_i=1\mid \bx_i) = \pi(\bx_i,\btheta) = \mathrm{logit}^{-1}\!\left(\bx_i^\T\btheta\right),
\end{equation*}
where $\mathrm{logit}^{-1}(u)=\exp(u)\{1+\exp(u)\}^{-1}$. In our applications, the auxiliary variables used in the logistic propensity score model are available for all units in the population frame, so the model is fitted on the full frame rather than on a reference probability sample. This is a direct simplification of Theorem~1 in \citet{CLW_2020}: under the MAR condition $\PP(\delta_i=1\mid \bx_i,y_i^\star)=\PP(\delta_i=1\mid \bx_i)$ and correct specification of the propensity score model, the estimator $\hat t^{\mathrm{IPW}}_{m\star}$ is asymptotically unbiased for $t_{m\star}$.

To estimate the sampling variability of the estimator \eqref{IPW_est}, we employ a pseudo-population bootstrap procedure adapted from \citet{LSW_2023}. Let $\hat w_i=1/\hat\pi_i$ for $i\in B$. The algorithm is:

\begin{enumerate}[label=(\alph*)]
\item Multiply the weights $\hat w_i$ by $N/\sum_{i\in B}\hat w_i$ to obtain $\tilde w_i$ satisfying $\sum_{i\in B}\tilde w_i=N$.

\item Randomly round $\tilde w_i$ to its ceiling with probability $\tilde w_i-\lfloor \tilde w_i \rfloor$ and to its floor otherwise to obtain $[\tilde w_i]$, with the requirement that $\sum_{i\in B}[\tilde w_i]=N$.

\item Create a pseudo-population by copying $[\tilde w_i]$ times each unit $i\in B$.

\item Draw a bootstrap non-probability sample $B^{*(r)}$ from the pseudo-population by Poisson sampling with inclusion probabilities $\hat\pi_i$.

\item Compute the bootstrap IPW estimators
\begin{equation}\label{eq:boot_rep}
\hat t^{\mathrm{IPW}*(r)}_{m\star} = \frac{N_m}{\hat N^{(r)}_{B_m}}\sum_{i\in B^{*(r)}_m}\frac{y_i^\star}{\hat\pi_i} \quad \text{with }\hat N^{(r)}_{B_m} = \sum_{i\in B^{*(r)}_m}\frac{1}{\hat\pi_i},
\end{equation}
where $B^{*(r)}_m=B^{*(r)}\cap U_m$ denotes the bootstrap sample for domain $m$.

\item Repeat steps (d)--(e) independently for $r=1,\ldots,R$ bootstrap replicates. 
\end{enumerate}
The bootstrap replicates \eqref{eq:boot_rep} will later be used to characterize the uncertainty of the auxiliary totals within the extended FH model.

\subsubsection{Incorporating the estimated auxiliary total: the Ybarra--Lohr formulation}\label{sec:YL}

We now extend the vector $\bz_m$ of exactly known auxiliary totals used in the classical FH formulation to $\bz_{m\star} = (z_{m0},z_{m1},\ldots,z_{mq},z_{m\star})^\T$ by adding the total $z_{m\star}=t_{m\star}$. For the Ybarra--Lohr extension, we also set $\hat z_{m\star} = \hat t^{\mathrm{IPW}}_{m\star}$ and accordingly write the estimated augmented auxiliary vector as $\hat\bz_{m\star} = (z_{m0},z_{m1},\ldots,z_{mq},\hat z_{m\star})^\T$.

Because the FH model is fitted to log-transformed domain totals of the response variable, we also work with the log-transformed auxiliary total $\log(\hat z_{m\star})$ and consider the MSE matrix of the vector $\log(\hat\bz_{m\star})$. In the Ybarra--Lohr framework, the uncertainty in the auxiliary information is summarized by
\begin{equation*}
\bC_{m\star} = \MSE\bigl(\log(\hat\bz_{m\star})\bigr),
\end{equation*}
which, in our setting, has zeros in all entries except for the last diagonal element: the components $z_{m0},\ldots,z_{mq}$ are treated as fixed and only $\hat z_{m\star}$ is error-prone. Under the MAR assumption and the conditions of \citet{CLW_2020}, we ignore potential bias in $\hat z_{m\star}$ and represent the measurement error solely through its variance on the log scale. Thus we take
\begin{equation*}
\bC_{m\star} = \mathrm{diag}\bigl(0,\ldots,0, C_{m\star}\bigr) \quad \text{with } C_{m\star} = \Var\bigl(\log (\hat z_{m\star})\bigr),
\end{equation*}
and estimate these quantities using the bootstrap replicates \eqref{eq:boot_rep} by defining
\begin{equation*}
\hat \bC_{m\star} = \mathrm{diag}\bigl(0,\ldots,0,\hat C_{m\star}\bigr),
\end{equation*}
where
\begin{equation}\label{eq:C_boot}
\hat C_{m\star} = \frac{1}{R-1}\sum_{r=1}^{R} \left[ \log\left(\hat t_{m\star}^{\mathrm{IPW}*(r)}\right) -\frac{1}{R}\sum_{r=1}^{R}\log\left(\hat t_{m\star}^{\mathrm{IPW}*(r)}\right) \right]^2,
\end{equation}
is the bootstrap estimator of $C_{m\star}$.

Let $\hat t_m^{\mathrm d}$ denote the H\'ajek, GREG, or any other traditional direct estimator of the domain total $t_m$, and define its log-transformed version $\hat t_m^{\mathrm d(\log)}=\log(\hat t_m^{\mathrm d})$. In the presence of the additional estimated auxiliary total, we follow \citet{YL_2008} and fit the FH-type model
\begin{equation}\label{eq:FH_YL_log}
\hat t_m^{\mathrm d(\log)} = \hat\bz_{m\star}^{\T}\bbeta + u_m + e_m, \quad m=1,\ldots, M,
\end{equation}
where $\bbeta=(\beta_0,\beta_1,\ldots,\beta_q,\beta_{q+1})^\T$ is the vector of regression coefficients, $u_m\overset{\text{ind}}{\sim}\mathcal{N}(0,\sigma_u^2)$ are area-specific random effects, and $e_m\overset{\text{ind}}{\sim}\mathcal{N}(0,\psi_m^{(\log)})$ are sampling errors. Here $\psi_m^{(\log)}$ denotes the sampling variance of $\hat t_m^{\mathrm d(\log)}$. Its design-based estimator $\hat\psi_m^{(\log)}$ is obtained in the usual way for probability samples, for example by the \citet{RWY_1992} or \citet{BBH_1994} bootstrap. The measurement-error FH formulation assumes that the sampling errors $e_m$ are independent of the measurement error in the auxiliary total. In the model \eqref{eq:FH_YL_log}, we use the estimated auxiliary vector $\hat\bz_{m\star}$ together with its measurement error structure, summarized by the matrix $\hat\bC_{m\star}$. 

Under this measurement error FH model, the EBLUP of the log-transformed domain total $t_m^{(\log)}$ can be written in the familiar shrinkage form
\begin{equation}\label{eq:EBLUP_YL_log}
\tilde t_m^{\mathrm{YL}(\log)} = \hat\gamma_m^{\mathrm{YL}}\hat t_m^{\mathrm d(\log)} + \bigl(1-\hat\gamma_m^{\mathrm{YL}}\bigr) \hat\bz_{m\star}^{\T}\hat\bbeta,
\end{equation}
where $\hat\bbeta$ is the estimator of $\bbeta$ obtained under the model \eqref{eq:FH_YL_log} and $\hat\gamma_m^{\mathrm{YL}}$ is the shrinkage factor. Owing to the simple structure of $\hat\bC_{m\star}$, the additional variance term $\hat\bbeta^{\T}\hat\bC_{m\star}\hat\bbeta$ appearing in the Ybarra--Lohr shrinkage factor reduces to $\hat\beta_{q+1}^2\hat C_{m\star}$, so that
\begin{equation*}
\hat\gamma_m^{\mathrm{YL}} = \frac{\hat\sigma_u^2 + \hat\beta_{q+1}^2\hat C_{m\star}}{\hat\sigma_u^2 + \hat\psi_m^{(\log)} + \hat\beta_{q+1}^2\hat C_{m\star}},
\end{equation*}
where $\hat C_{m\star}$ is the variance estimator given by \eqref{eq:C_boot} and $\hat\sigma_u^2$ is an estimator of $\sigma_u^2$.

The regression coefficients $\bbeta$ and the variance component $\sigma_u^2$ are estimated by adapting the weighted normal equations of \citet{YL_2008} to the present setting, which involves a single error-prone auxiliary total. In essence, the usual FH normal equations are modified by replacing the unknown auxiliary total by its estimator $\hat z_{m\star}$ and by subtracting $\hat\bC_{m\star}$ from the corresponding cross-product matrix, so that the working variances in the weights reflect the total variability $\sigma_u^2+\psi_m^{(\log)}+\beta_{q+1}^2 C_{m\star}$. This leads to an iteratively reweighted least squares procedure, in which $\bbeta$ and $\sigma_u^2$ are updated until convergence. The resulting estimation algorithm is implemented for the measurement error FH model in the \texttt{emdi} package for \textsf{R} \citep[see][]{HKSSS_2023}. We do not reproduce the full estimating equations here and refer the reader to \citet{YL_2008} for explicit formulae.

For the predictor \eqref{eq:EBLUP_YL_log}, we approximate its MSE by using a jackknife-type estimator developed within the general theory for empirical best prediction in mixed models \citep{JLW_2002}. In practice, we rely on the implementation for the measurement error FH model provided in the \texttt{emdi} package for \textsf{R} \citep{HKSSS_2023}, which constructs delete-one-area replicates of \eqref{eq:EBLUP_YL_log} and combines them according to the unified jackknife framework of \citet{JLW_2002}. The resulting estimator of $\MSE(\tilde t_m^{\mathrm{YL}(\log)})$ is then used in exactly the same way as in Section~\ref{sec:FH} when back-transforming from the log scale to the original scale. Thus, we obtain the predictor on the original scale, denoted by $\hat t_m^{\mathrm{YL}}$, and approximate its MSE by applying the delta method to the jackknife-based MSE on the log scale.

\section{Applications to official statistics surveys}\label{sec:results}

\subsection{Data and design}\label{sec:data}

We apply the proposed estimators to three enterprise surveys conducted by the State Data Agency (Statistics Lithuania). In each application, a probability sample $A$ is selected by stratified simple random sampling without replacement, and it is complemented by an external non-probability sample $B$, which is typically larger than $A$ and can be linked at the enterprise level. The sample $B$ contains a proxy variable $y^\star$ related to the study variable $y$, and the overlap $A\cap B$, with the design weights inherited from the probability sample, is used to specify a working measurement-error model linking $y$ to $(y^\star,\bx)$. Here, we use a vector of completely known auxiliary variables $\bx_i=(1,x_{i1},\ldots,x_{ip})^\T$; in the applications below, we take $p=1$, that is, $\bx_i=(1,x_{i1})^\T$.

\begin{enumerate}[topsep=3pt, itemsep=2pt]
\item \textit{Monthly turnover.} In the statistical survey on service enterprise activities, based on sample $A$, the study variable $y$ is the monthly turnover. The goal is to estimate its totals for domains defined by 4-digit groups of the Statistical Classification of Economic Activities in the European Community (NACE). Let the GREG estimator be the initial direct estimator, supported by the completely known auxiliary annual turnover from the previous year, denoted by $x_{i1}$. The proxy variable $y^\star$ is a monthly turnover derived from VAT declaration data. Given the close alignment between the survey- and VAT-based turnover, a simple linear measurement-error model is an appropriate link between $y$ and $(y^\star,\bx)$ on the overlap of $A$ and $B$.

\item \textit{Annual investment in tangible assets.} In the statistical survey on investment, the study variable $y$ is the annual investment in tangible assets. The objective is to estimate the totals of $y$ for domains defined by 2-digit NACE activity groups. The survey currently uses the Horvitz--Thompson estimator based on sample $A$ only. A slightly larger non-probability sample $B$ is available from an alternative structural business statistics (SBS) survey also conducted by the State Data Agency. This is a same-year cut-off sample that provides similar investment values for enterprises exceeding specified size thresholds, which we treat as the proxy $y^\star$. The latter SBS survey also provides the previous year’s investment $x_{i1}$, which is completely known in $U$. Since many enterprises report zeros while positive values are continuous and right-skewed, we adopt a two-part (hurdle) measurement-error model, consisting of a logistic component for the probability of positive investment and a log-linear regression component for the positive values.

\item \textit{Quarterly job vacancies.} The values of the study variable $y$ are measurements of job vacancies at the end of the quarter in the sample $A$ of the statistical survey on earnings. The aim is to estimate the totals of this count variable $y$ by municipality, while the stratification in the sampling design is based on NACE activity and enterprise size groups. Unlike the other two applications, municipalities are not explicit design domains, so the number of sampled enterprises can be small in some municipalities. Therefore, the Hájek estimator \eqref{eq:dir_H} was used in producing official estimates. The non-probability sample $B$ provides web-scraped counts of OJA, which we use as the proxy $y^\star$. The monthly number of employees $x_{i1}$ is known for all enterprises and serves as a completely known auxiliary variable. Since vacancy counts are sparse with many zeros, we use a nonparametric $k$NN measurement-error model.
\end{enumerate}

The linkage is deterministic in the first two applications. Survey records, VAT declarations and the SBS cut-off sample use the same official legal entity code, which is common to Lithuanian enterprise surveys and administrative sources. In the OJA application, linkage is based on standardized enterprise names. Names in the register and in the web-scraped data are cleaned by the same rules, industry-specific dictionaries of common name components are used to remove non-informative words, and exact matching is then applied to the remaining standardized names. Ambiguous one-to-many matches are resolved by a random-forest classifier. In practical production, the resulting linked OJA counts are considered sufficiently reliable for use as auxiliary information in the model-calibration step.

\subsection{Estimators, precision measures and working models}\label{sec:app_estim}

Let us summarize the estimators reported in the three applications, the precision measures used to evaluate them, and the working models required for implementing the model-calibration approach. For each domain $U_m$, we consider the direct estimator $\hat t_m^{\mathrm d}$ used in the corresponding survey (GREG for monthly turnover, Horvitz--Thompson for annual investment, and Hájek for quarterly job vacancies), the model-calibration estimator $\hat t_m^{\mathrm{MC}}$ defined in \eqref{eq:mc_domain}, the EBLUP $\hat t_m^{\mathrm{EBLUP}}$ given by \eqref{eq:EBLUP_back}, and the Ybarra--Lohr predictor $\hat t_m^{\mathrm{YL}}$ constructed in Section~\ref{sec:YL}. For brevity, we denote the direct estimators by GREG, HT, and H, and the additional estimators by MC, EBLUP, and YL in the tables and figures. In all cases, we use the variance and MSE estimators introduced in Section~\ref{sec:methodology}.

To summarize precision at the domain level, we report the estimated relative root mean squared error (RRMSE)
\begin{equation*}
\mathrm{RRMSE}(\hat t_m)=\frac{\sqrt{\widehat{\mathrm{MSE}}(\hat t_m)}}{\hat t_m},
\end{equation*}
where $\hat t_m$ is any estimate of the domain total $t_m$. For nearly unbiased direct and other estimators, this quantity is close to the usual coefficient of variation. Let us classify domain estimates as ``good'' if RRMSE is below $0.167$, ``sufficient'' if it lies between $0.167$ and $0.333$, and ``unreliable'' otherwise.

To implement the model-calibration constraint \eqref{eq:mc_constr3}, we specify working models in \eqref{eq:me_model} and compute predictions $\hat y_i=\hat g(y_i^\star,\bx_i)$ for units in $B$, using design-weighted model fits obtained from the overlap $A\cap B$. In all three applications we take $\bx_i=(1,x_{i1})^\T$. The working models used in the applications are specified as follows.

\begin{enumerate}[topsep=3pt, itemsep=2pt]
\item \textit{Monthly turnover:} we use a linear working model
\begin{equation*}
g_{\mathrm{turn}}(y_i^\star,\bx_i;\bbeta,\beta_\star)=\bbeta^\T \bx_i+\beta_\star y_i^\star.
\end{equation*}
The parameters are estimated by design-weighted least squares over $A\cap B$,
\begin{equation*}
(\hat\bbeta,\hat\beta_\star)=\argmin_{\bbeta,\beta_\star} \sum_{i\in A\cap B} d_i\left(y_i-\bbeta^\T\bx_i-\beta_\star y_i^\star\right)^2,
\end{equation*}
and $\hat y_i=\hat g_{\mathrm{turn}}(y_i^\star,\bx_i)$ is computed for all $i\in B$.

\item \textit{Annual investment in tangible assets:} we use a two-part working model. Let $\br_i=(1,\log(1+y_i^\star),\log(1+x_{i1}))^\T$, using logarithms for these nonnegative, right-skewed business variables. First, the probability of a positive value is modeled as
\begin{equation*}
p_i(\bgamma)=\PP(y_i>0\mid y_i^\star,\bx_i)=\mathrm{logit}^{-1}\!\left(\bgamma^\T\br_i\right),
\end{equation*}
where $\bgamma$ is estimated from the design-weighted logistic score equation
\begin{equation*}
\sum_{i\in A\cap B} d_i\,\br_i\{I(y_i>0)-p_i(\hat\bgamma)\}=0.
\end{equation*}
Second, for positive values,
\begin{equation*}
\log(1+y_i)=\balpha^\T\br_i+\varepsilon_i,\quad y_i>0,
\end{equation*}
with $\balpha$ estimated by design-weighted least squares over $\{i\in A\cap B:y_i>0\}$. The prediction is
\begin{equation*}
\hat y_i=\hat p_i\left[\exp\left(\hat\balpha^\T\br_i+\hat\sigma_+^2/2\right)-1\right], \quad i\in B,
\end{equation*}
where $\hat p_i=p_i(\hat\bgamma)$ and $\hat\sigma_+^2$ is the design-weighted residual variance from the positive-part model.

\item \textit{Quarterly job vacancies:} we use a nonparametric $k$NN working model. Let $D_{ij}$ be the Gower distance between $(y_i^\star,x_{i1})$ and $(y_j^\star,x_{j1})$, and let $N_k(i)\subset A\cap B$ be the set of $k$ nearest neighbors of unit $i\in B$. In this application we use $k=5$ and inverse-distance weights
\begin{equation*}
a_{ij}=(D_{ij}+10^{-8})^{-1};
\end{equation*}
other kernel weights could be used analogously. The design-weighted $k$NN prediction is
\begin{equation*}
\hat g_{\mathrm{vac}}(y_i^\star,\bx_i)=\frac{\sum_{j\in N_k(i)} d_j a_{ij} y_j}{\sum_{j\in N_k(i)} d_j a_{ij}},
\end{equation*}
and we set $\hat y_i=\hat g_{\mathrm{vac}}(y_i^\star,\bx_i)$ for all $i\in B$.
\end{enumerate}

The three applications followed a common operational strategy: domains were screened using their sample sizes and the representation of the proxy-observed and proxy-missing post-strata in the probability sample (a full-rank calibration matrix also requires at least two distinct fitted values within $A_m\cap B_m$), and linear calibration was performed with positive lower bounds on the weights. Where MC was not used, the corresponding direct estimate and its design-based variance were retained as FH inputs where applicable; exact take-all totals were retained without FH prediction. These empirical safeguards were distinct from the simulation-specific numerical feasibility criteria in Section~S.2.3 of the supplementary material.

Table~\ref{tab:calibration_stability} summarizes the observed area sample sizes and the operational use of the model-calibration estimator.

\begin{table}[H]
    \centering
    \caption{Area sample sizes and use of the model-calibration estimator in the three applications.}
    \label{tab:calibration_stability}
    {\small\setlength{\tabcolsep}{4pt}
    \resizebox{\linewidth}{!}{%
    \begin{tabular}{@{}lcccccc@{}}
        \toprule
        Application & \begin{tabular}{c}Periods;\\areas per period\end{tabular}
        & $|A_m|$ & $|A_m\cap B_m|$ & $|A_m\setminus B_m|$
        & MC used & MC not used \\
        \midrule
        Monthly turnover        & 12; 71 & 35 [26, 54] & 29 [18, 46] & 4 [2, 9]    & 58.5 [58, 59] & 12.5 [12, 13] \\
        Annual investment       & 1; 80  & 72 [42, 143] & 19 [6, 36]  & 52 [36, 106] & 67 & 13 \\
        Quarterly job vacancies & 4; 60  & 44 [32, 71] & 19.5 [13, 34] & 26 [17, 38] & 47.5 [43, 51] & 12.5 [9, 17] \\
        \bottomrule
    \end{tabular}}}
    \tabnote{Sample-size entries are medians [first quartile, third quartile] over all area-period combinations. For applications with several periods, the numbers of areas where model-calibration (MC) was and was not used are period means [minimum, maximum]; for annual investment they are exact counts. When MC was not used, the direct estimate and its design-based variance served as first-stage Fay--Herriot model inputs where applicable; exact take-all domains and domains not entering the Fay--Herriot fit retained the direct estimator.}
\end{table}

To assess the log-normal approximation in \eqref{eq:EBLUP_back}, we applied the Shapiro--Wilk test to standardized marginal residuals from the fitted FH models, computed from the log-transformed first-stage estimates and their corresponding sampling variances, separately for each period. Normality was not rejected at the 5\% level in 4 of the 12 turnover months and 2 of the 4 job-vacancy quarters; for annual investment, the $p$-value was 0.024. The maximum absolute standardized residuals ranged from 2.57 to 4.96 for turnover and from 2.79 to 4.96 for job vacancies, and equalled 3.85 for investment. The period-specific medians of the multiplicative log-normal correction factor $\exp\{0.5\,\widehat{\mathrm{MSE}}(\tilde t_m^{(\log)})\}$ in \eqref{eq:EBLUP_back} remained close to one, ranging from 1.000 to 1.002 for turnover and from 1.010 to 1.013 for job vacancies, and equalling 1.006 for investment. The respective maxima of this factor over all areas and periods were 1.030, 1.136, and 1.064. Although these diagnostics indicate some tail departures, the typical corrections were small, and we retained the log-normal correction.

\subsection{Monthly turnover}\label{sec:turnover}

In this application, we estimate monthly turnover totals for $M=71$ domains defined by 4-digit NACE activity groups in each of $T=12$ months, using the four estimators: the direct GREG estimator, the model-calibration estimator $\hat t_m^{\mathrm{MC}}$, the EBLUP $\hat t_m^{\mathrm{EBLUP}}$, and the Ybarra--Lohr predictor $\hat t_m^{\mathrm{YL}}$. To isolate the effect of model calibration, the summaries below are restricted to the domains where the MC estimator was used, averaging 58.5 domains per month (58--59). The remaining domains are included in the all-area operational assessment in Section~\ref{sec:summary}. The average relative size is $|B|/|A|\approx 6.2$ across months.

\begin{figure}[H]
    \centering
    \caption{Monthly turnover example: summary of domain-level relative root mean squared error across months for the domains where the model-calibration estimator was used.}
    \label{fig:turnover_rrmse}
    \includegraphics[width = \linewidth]{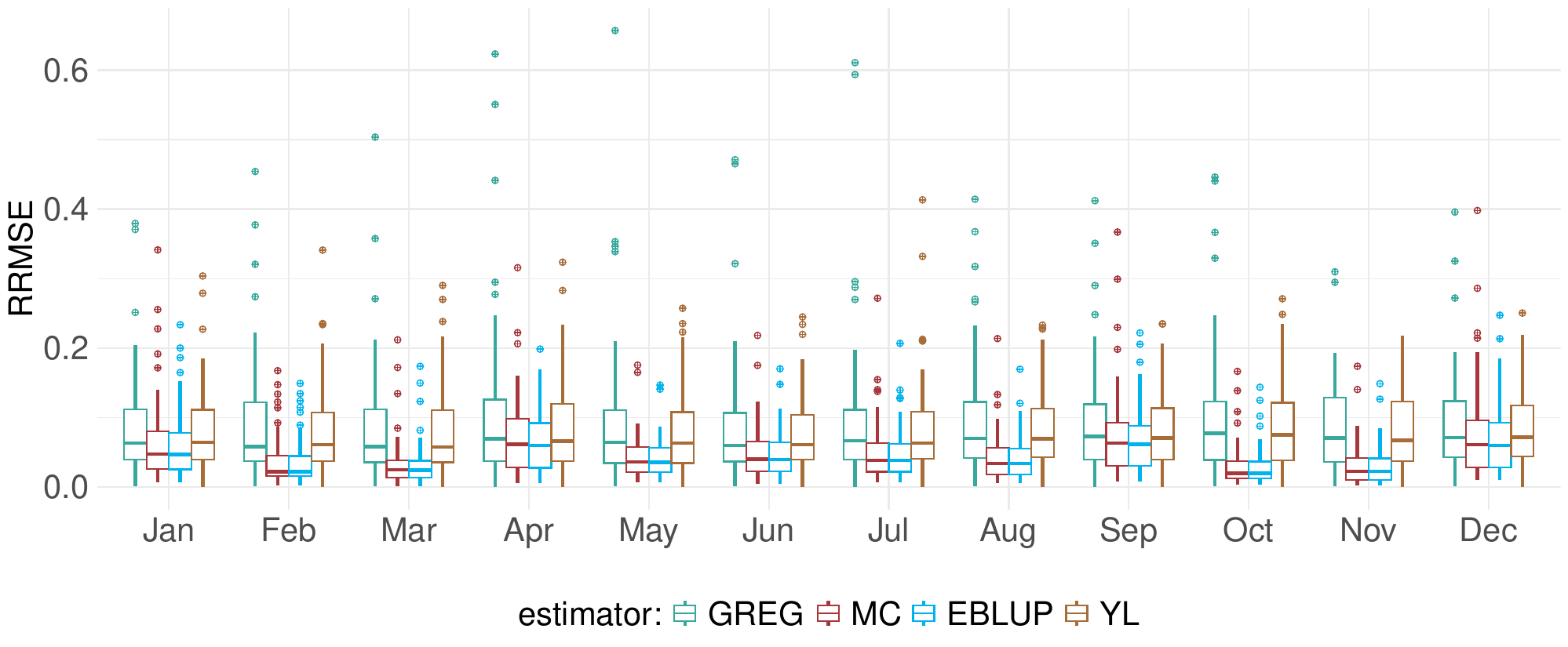}
    \fignote{RRMSE = relative root mean squared error; GREG = generalized regression estimator; MC = model-calibration estimator; EBLUP = empirical best linear unbiased predictor; YL = Ybarra--Lohr predictor.}
\end{figure}

To present the results in a compact way, Table~\ref{tab:turnover_quality} summarizes domain reliability classifications based on RRMSE (good / sufficient / unreliable) for the domains where the model-calibration estimator was used, averaged over the 12 months considered. Figure~\ref{fig:turnover_rrmse} summarizes how the corresponding distribution of domain-level RRMSE evolves across months for the four estimators.

\begin{table}[H]
    \centering
    \caption{Monthly turnover example: average (over 12 months) number of 4-digit activity groups according to the Statistical Classification of Economic Activities in the European Community in which the model-calibration estimator was used, classified by relative root mean squared error quality categories for each estimator.}
    \label{tab:turnover_quality}
    {\setlength{\tabcolsep}{6pt}
    \begin{tabular}{@{}l*{3}{>{\centering\arraybackslash}p{2.3cm}}@{}}
        \toprule
        Estimator & Good & Sufficient & Unreliable \\
        \midrule
        GREG  & 50.8 & 5.6 & 2.1 \\
        MC    & 56.3 & 1.9 & 0.2 \\
        EBLUP & 57.2 & 1.3 & 0.0 \\
        YL    & 52.7 & 5.7 & 0.2 \\
        \bottomrule
    \end{tabular}}
    \tabnote{GREG = generalized regression estimator; MC = model-calibration estimator; EBLUP = empirical best linear unbiased predictor; YL = Ybarra--Lohr predictor.}
\end{table}

Table~\ref{tab:turnover_quality} shows that the direct GREG estimator already performs reasonably well, with 50.8 of the 58.5 domains where model calibration was used classified as good on average, but it leaves 2.1 domains per month in the unreliable category (RRMSE $>0.333$). Incorporating the VAT-based proxy through model calibration increases the average number of good domains to 56.3 and reduces the unreliable count to 0.2. Figure~\ref{fig:turnover_rrmse} shows a clear downward shift in the MC distribution across months, particularly in its upper part. The subsequent FH step provides a smaller additional gain: EBLUP yields 57.2 good domains and no unreliable domains on average, with a slightly shorter upper tail than MC. The Ybarra--Lohr predictor also improves the extreme tail relative to GREG, but its central RRMSE distribution remains much closer to the direct estimator than to MC or EBLUP.

\subsection{Annual investment in tangible assets}\label{sec:investment}

In this application, we estimate annual investment totals for $M=80$ domains defined by 2-digit NACE activity groups in 2023, using the four estimators: the direct HT estimator, the model-calibration estimator $\hat t_m^{\mathrm{MC}}$, the EBLUP $\hat t_m^{\mathrm{EBLUP}}$, and the Ybarra--Lohr predictor $\hat t_m^{\mathrm{YL}}$. To isolate the effect of model calibration, the comparison covers the 67 domains where the MC estimator was used. The remaining 13 domains are included in the all-area operational assessment in Section~\ref{sec:summary}. The relative size is $|B|/|A|\approx 1.1$.

\begin{figure}[H]
    \centering
    \caption{Annual investment example: summary of domain-level relative root mean squared error for the domains where the model-calibration estimator was used in 2023.}
    \label{fig:investment_rrmse}
    \includegraphics[width = 0.75\linewidth]{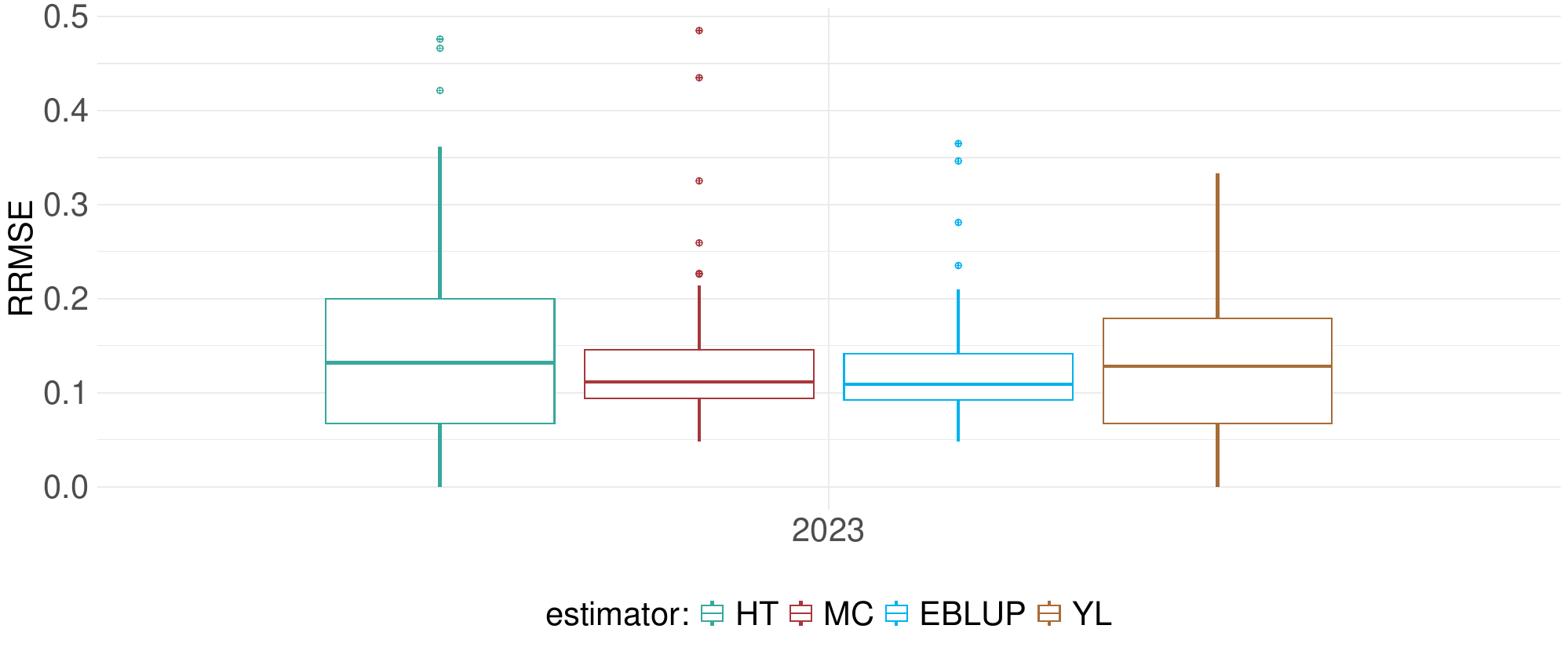}
    \fignote{RRMSE = relative root mean squared error; HT = Horvitz--Thompson estimator; MC = model-calibration estimator; EBLUP = empirical best linear unbiased predictor; YL = Ybarra--Lohr predictor.}
\end{figure}

To present the results in a compact way, Table~\ref{tab:investment_quality} summarizes domain reliability classifications based on RRMSE (good / sufficient / unreliable) among these 67 domains. Figure~\ref{fig:investment_rrmse} summarizes the corresponding distribution of domain-level RRMSE for the four estimators.

\begin{table}[H]
    \centering
    \caption{Annual investment example: number of 2-digit activity groups according to the Statistical Classification of Economic Activities in the European Community in which the model-calibration estimator was used, classified by relative root mean squared error quality categories for each estimator in 2023.}
    \label{tab:investment_quality}
    {\setlength{\tabcolsep}{6pt}
    \begin{tabular}{@{}l*{3}{>{\centering\arraybackslash}p{2.3cm}}@{}}
        \toprule
        Estimator & Good & Sufficient & Unreliable \\
        \midrule
        HT        & 37 & 26 & 4 \\
        MC        & 54 & 11 & 2 \\
        EBLUP     & 56 &  9 & 2 \\
        YL        & 43 & 23 & 1 \\
        \bottomrule
    \end{tabular}}
    \tabnote{HT = Horvitz--Thompson estimator; MC = model-calibration estimator; EBLUP = empirical best linear unbiased predictor; YL = Ybarra--Lohr predictor.}
\end{table}

Compared with the turnover application, this setting is more challenging because investment values are strongly zero-inflated and the non-probability sample is only slightly larger than the probability sample ($|B|/|A|\approx 1.1$). Among these 67 domains, Table~\ref{tab:investment_quality} shows that MC increases the number classified as good from 37 to 54, reduces the sufficient category from 26 to 11, and halves the unreliable count from 4 to 2. The FH step provides a smaller additional improvement, yielding 56 good domains while retaining 2 unreliable domains. Figure~\ref{fig:investment_rrmse} shows that MC and EBLUP lower the center of the distribution and contract its upper part. The Ybarra--Lohr predictor has a center closer to HT and reduces the number of unreliable domains from 4 to 1.

\subsection{Quarterly job vacancies}\label{sec:vacancies}

In this application, we estimate job vacancy totals for $M=60$ municipalities in each of $T=4$ quarters, using the four estimators: the direct Hájek estimator, the model-calibration estimator $\hat t_m^{\mathrm{MC}}$, the EBLUP $\hat t_m^{\mathrm{EBLUP}}$, and the Ybarra--Lohr predictor $\hat t_m^{\mathrm{YL}}$. To isolate the effect of model calibration, the summaries below are restricted to the municipalities where the MC estimator was used, averaging 47.5 municipalities per quarter (43--51). The remaining municipalities are included in the all-area operational assessment in Section~\ref{sec:summary}. The average relative size is $|B|/|A|\approx 2.1$ across quarters.

\begin{figure}[H]
    \centering
    \caption{Quarterly job vacancies example: summary of domain-level relative root mean squared error across quarters for the municipalities where the model-calibration estimator was used.}
    \label{fig:vac_rrmse}
    \includegraphics[width = 0.8\linewidth]{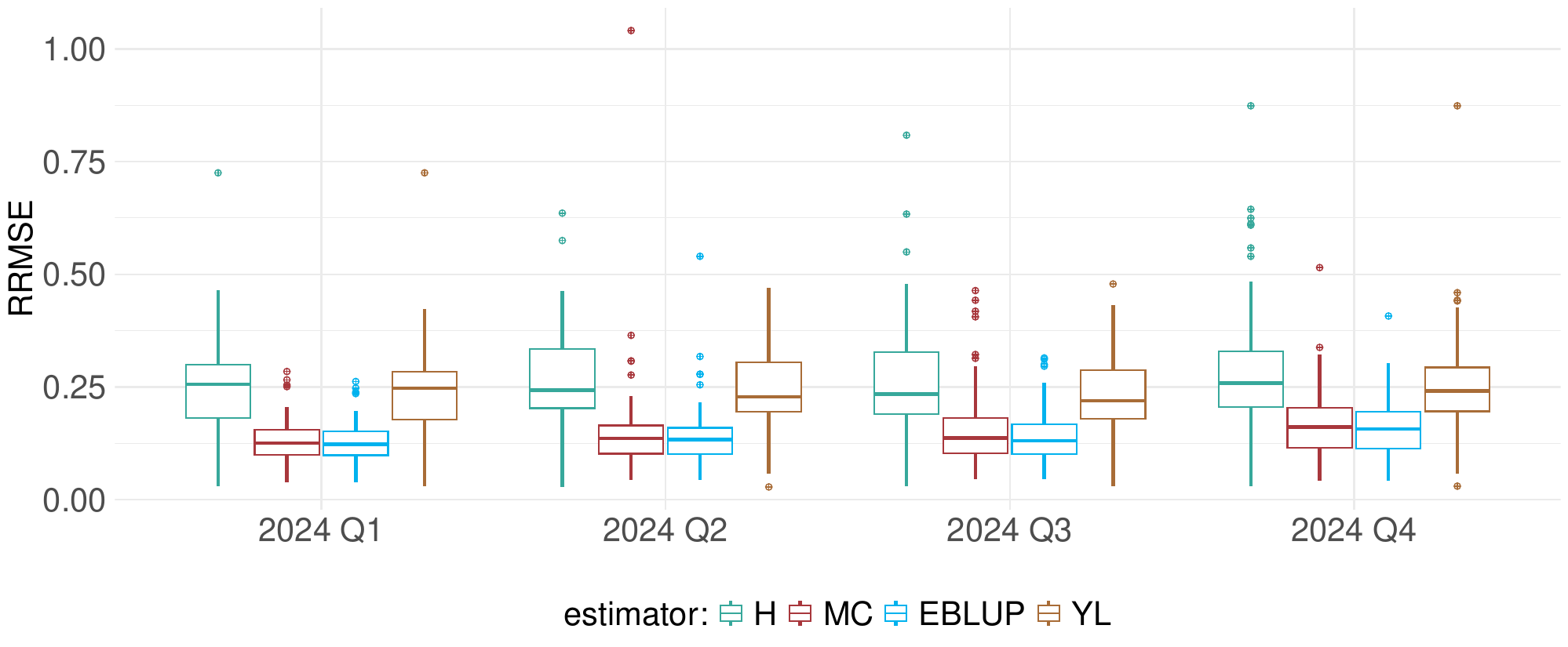}
    \fignote{RRMSE = relative root mean squared error; H = H\'ajek estimator; MC = model-calibration estimator; EBLUP = empirical best linear unbiased predictor; YL = Ybarra--Lohr predictor.}
\end{figure}

To present the results in a compact way, Table~\ref{tab:vac_quality} summarizes domain reliability classifications based on RRMSE (good / sufficient / unreliable) for the municipalities where the model-calibration estimator was used, averaged over the 4 quarters considered. Figure~\ref{fig:vac_rrmse} summarizes how the corresponding distribution of domain-level RRMSE evolves across quarters for the four estimators.

\begin{table}[H]
    \centering
    \caption{Quarterly job vacancies example: average (over 4 quarters) number of municipalities in which the model-calibration estimator was used, classified by relative root mean squared error quality categories for each estimator.}
    \label{tab:vac_quality}
    {\setlength{\tabcolsep}{6pt}
    \begin{tabular}{@{}l*{3}{>{\centering\arraybackslash}p{2.3cm}}@{}}
        \toprule
        Estimator & Good & Sufficient & Unreliable \\
        \midrule
        H         &  6.2 & 30.5 & 10.8 \\
        MC        & 32.8 & 12.8 &  2.0 \\
        EBLUP     & 34.8 & 12.2 &  0.5 \\
        YL        &  8.2 & 32.0 &  7.2 \\
        \bottomrule
    \end{tabular}}
    \tabnote{H = H\'ajek estimator; MC = model-calibration estimator; EBLUP = empirical best linear unbiased predictor; YL = Ybarra--Lohr predictor.}
\end{table}

The municipal job-vacancy application shows the most pronounced change in the reliability classification. As summarized in Table~\ref{tab:vac_quality}, the direct H\'ajek estimator yields only 6.2 good municipalities on average and leaves 10.8 in the unreliable category. After integrating OJA information through the $k$NN-based model-calibration step, the average number of good municipalities increases to 32.8 and the unreliable count falls to 2.0. Figure~\ref{fig:vac_rrmse} shows substantially lower MC medians and upper quartiles in all four quarters. The FH EBLUP provides a further gain, increasing the average number of good municipalities to 34.8 and reducing the unreliable count to 0.5. By contrast, the Ybarra--Lohr predictor produces a more modest improvement, with 8.2 good and 7.2 unreliable municipalities on average.

\subsection{Comparative summary across the three examples}\label{sec:summary}

We conclude the application section by comparing two complementary evaluations. The MC-supported-area columns of Table~\ref{tab:app_summary} restrict the comparison to areas where the model-calibration estimator was used and therefore isolate the effect of MC. The all-area columns evaluate the operational procedure across every area: MC is used where available and the direct estimator otherwise; the FH fit uses the corresponding MC or direct first-stage estimate and variance, while exact take-all domains and domains not entering the FH fit retain the direct estimate. For the Ybarra--Lohr benchmark, the direct estimator is used where the proxy total required by that method is unavailable. For monthly turnover and quarterly job vacancies, the reported quantities are averages of period-specific quantiles; for annual investment, they refer to 2023.

\begin{table}[H]
    \centering
    \caption{Comparative summary of domain-level relative root mean squared error in the three applications, for areas where the model-calibration estimator was used and for all areas under the operational procedure.}
    \label{tab:app_summary}
    {\setlength{\tabcolsep}{5pt}
    \begin{tabular}{@{}l*{4}{>{\centering\arraybackslash}p{1.9cm}}@{}}
        \toprule
        & \multicolumn{2}{c}{MC-supported areas} & \multicolumn{2}{c}{All areas} \\
        \cmidrule(lr){2-3}\cmidrule(l){4-5}
        Estimator & Median & P90 & Median & P90 \\
        \midrule

        \multicolumn{5}{@{}l@{}}{\textit{Monthly turnover (4-digit NACE), 12 months}} \\
        GREG  & 0.067 & 0.186 & 0.053 & 0.171 \\
        MC    & 0.039 & 0.099 & 0.032 & 0.096 \\
        EBLUP & 0.039 & 0.093 & 0.032 & 0.090 \\
        YL    & 0.066 & 0.165 & 0.054 & 0.154 \\
        \addlinespace[2pt]

        \multicolumn{5}{@{}l@{}}{\textit{Annual investment (2-digit NACE), 2023}} \\
        HT    & 0.132 & 0.299 & 0.132 & 0.333 \\
        MC    & 0.111 & 0.208 & 0.112 & 0.266 \\
        EBLUP & 0.109 & 0.195 & 0.110 & 0.240 \\
        YL    & 0.128 & 0.244 & 0.127 & 0.263 \\
        \addlinespace[2pt]

        \multicolumn{5}{@{}l@{}}{\textit{Quarterly job vacancies (municipalities), 4 quarters}} \\
        H     & 0.248 & 0.441 & 0.274 & 0.510 \\
        MC    & 0.140 & 0.267 & 0.152 & 0.451 \\
        EBLUP & 0.136 & 0.239 & 0.146 & 0.292 \\
        YL    & 0.234 & 0.374 & 0.254 & 0.419 \\
        \bottomrule
    \end{tabular}}
    \tabnote{P90 = the ninth decile of domain-level RRMSE; NACE = Statistical Classification of Economic Activities in the European Community; GREG = generalized regression estimator; HT = Horvitz--Thompson estimator; H = H\'ajek estimator; MC = model-calibration estimator; EBLUP = empirical best linear unbiased predictor; YL = Ybarra--Lohr predictor. The operational fallback rules are described in Section~\ref{sec:FH}.}
\end{table}

Within MC-supported areas, Table~\ref{tab:app_summary} confirms that most of the precision gain is achieved at the model-calibration stage, with the FH step generally providing a smaller additional tightening. For monthly turnover, MC reduces the median RRMSE from 0.067 to 0.039 and P90 from 0.186 to 0.099; EBLUP further reduces P90 to 0.093. For annual investment, the corresponding changes are from 0.132 to 0.111 and from 0.299 to 0.208, with EBLUP reducing them further to 0.109 and 0.195. For quarterly job vacancies, MC reduces the median from 0.248 to 0.140 and P90 from 0.441 to 0.267; EBLUP further lowers these values to 0.136 and 0.239.

The all-area assessment shows that these gains are not an artifact of restricting attention to MC-supported areas. For turnover, the direct-estimator median and P90 of 0.053 and 0.171 decline to 0.032 and 0.096 under MC with fallback, and to 0.032 and 0.090 under EBLUP. For investment, the corresponding sequence is 0.132/0.333, 0.112/0.266, and 0.110/0.240. For job vacancies, it is 0.274/0.510, 0.152/0.451, and 0.146/0.292. The more modest MC reduction in the all-area P90 for job vacancies reflects the concentration of difficult cases among fallback municipalities; using their direct estimates and variances in the FH model substantially reduces this upper tail.

Across applications, the largest reductions occur in turnover and job vacancies, while the gains for investment are more moderate. This ordering is not determined by the relative size $|B|/|A|$ alone: the job-vacancy application has a smaller relative non-probability sample than turnover but a comparable reduction in the median RRMSE. The effectiveness of the calibration constraints therefore depends jointly on the amount of proxy information, the strength of the unit-level working relationship, and the baseline precision of the direct estimators.

The Ybarra--Lohr predictor shows smaller and more tail-focused gains in both evaluations. Its median RRMSE remains close to that of the direct estimator, while the upper tail improves more visibly. By comparison, model calibration shifts both the center and the upper tail of the distribution. The FH contribution is generally smaller within MC-supported areas, but can become more important in the all-area upper tail when the FH model uses direct first-stage estimates and their design-based variances for fallback areas.

\section{Design-based simulation study}\label{sec:simulation}

The empirical applications cannot separate the effects of misspecifying the working prediction function, reduced overlap between the probability and non-probability samples, or violation of the MAR assumption used by the YL benchmark. We therefore conducted a focused design-based simulation. A single finite population was held fixed throughout the experiment. For each combination of selection mechanism and target coverage, one realization of the non-probability sample $B$ was generated, held fixed, and shared by the correctly specified and misspecified working-function scenarios, while repeated probability samples $A$ were drawn under the known sampling design. The experiment therefore evaluates repeated-sampling properties under the probability design in a manner consistent with the inferential framework used in Sections~2.1--2.5.

\subsection{Simulation design}\label{sec:sim_design}

The finite population contained $M=60$ domains and $21\,600$ units. Domain sizes increased from 220 to 500 units. Independently generated domain effects $a_m,u_m\sim\mathcal{N}(0,1)$ and unit errors $\xi_{mi},v_{mi}\sim\mathcal{N}(0,1)$ and $\varepsilon_{mi}\sim\mathcal{N}(0,4^2)$ were used to construct
\begin{align}
 x_{mi} &= a_m+\xi_{mi}, \label{eq:sim_x_final}\\
 y_{mi}^{\star} &= 5+1.2x_{mi}+0.6a_m+v_{mi}, \label{eq:sim_proxy_final}\\
 y_{mi} &= 50+3x_{mi}+7y_{mi}^{\star}+1.5(y_{mi}^{\star}-5)^2+4u_m+\varepsilon_{mi}. \label{eq:sim_y_final}
\end{align}
The 60 domains coincided with the sampling strata. A simple random sample without replacement of size $n_m=30$ was drawn independently within each domain/stratum, with design weight $d_{mi}=N_m/n_m$. The direct estimator was a domain GREG estimator calibrated to the known domain size and total of $x$. The fully observed area-level covariates used in the FH models were $\bz_m=(1,\log N_m,\bar x_m)^\T$, where $\bar x_m=N_m^{-1}\sum_{i\in U_m}x_{mi}$.

The non-probability sample was generated from
\begin{equation}\label{eq:sim_B_final}
 \PP(\delta_{mi}=1)
 =\operatorname{logit}^{-1}\!\left\{
 \alpha_{\rho}+0.7x_{mi}-0.4\bar x_m+\lambda r_{mi}
 \right\},
 \qquad
 r_{mi}=\frac{v_{mi}+\varepsilon_{mi}/4}{\sqrt{2}},
\end{equation}
where $\alpha_{\rho}$ was chosen to attain target coverage $\rho\in\{0.20,0.40,0.70\}$. Under MAR, $\lambda=0$ and the propensity score model fitted from the full population frame was correctly specified. Under non-MAR, $\lambda=1$, whereas the fitted propensity score model continued to use only the frame variables $x_{mi}$ and $\bar x_m$. The three coverage levels produced median expected overlap sizes $|A_m\cap B_m|$ of approximately 6, 12 and 21 units per domain.

The correctly specified working function included the quadratic term in \eqref{eq:sim_y_final},
\begin{equation}\label{eq:sim_g_correct_final}
 g_{\mathrm C}(y^{\star},x)
 =\theta_0+\theta_1x+\theta_2y^{\star}+\theta_3(y^{\star}-5)^2,
\end{equation}
whereas the misspecified version omitted it,
\begin{equation}\label{eq:sim_g_miss_final}
 g_{\mathrm M}(y^{\star},x)
 =\theta_0+\theta_1x+\theta_2y^{\star}.
\end{equation}
Both functions were re-estimated from $A\cap B$ by design-weighted least squares in every probability-sample replication. The complete $2\times2\times3$ factorial experiment comprised 12 scenarios. Table~\ref{tab:sim_scenarios_final} lists the six core scenarios retained for the main text; the remaining scenarios are reported in the supplementary material.

\begin{table}[H]
\centering
\small
\caption{Core scenarios in the design-based simulation.}
\label{tab:sim_scenarios_final}
\begin{tabularx}{\textwidth}{@{}c l l c X@{}}
\toprule
Scenario & Working function & Selection into $B$ & Target coverage & Purpose \\
\midrule
S1 & Correct & MAR & 0.40 & Baseline \\
S2 & Misspecified & MAR & 0.40 & Effect of misspecifying $g$ \\
S3 & Correct & Non-MAR & 0.40 & Effect of violating MAR \\
S4 & Misspecified & Non-MAR & 0.40 & Combined adverse case \\
S5 & Correct & MAR & 0.70 & High-overlap benchmark \\
S6 & Correct & MAR & 0.20 & Low-overlap stress case \\
\bottomrule
\end{tabularx}
\tabnote{MAR = missing-at-random.}
\end{table}

Four estimators were compared: the direct GREG estimator, the model-calibration (MC) estimator with the operational direct-estimator fallback, the corresponding EBLUP, and the Ybarra--Lohr (YL) predictor. The general feasibility and direct-estimator fallback principles of Sections~2.2 and 2.4 were implemented using the simulation-specific numerical criteria listed in Section~S.2.3 of the supplementary material. For the YL predictor, the propensity score model was fitted once from the full population frame for each realized $B$, and the domain proxy totals were obtained by the normalized IPW estimator of Section~2.5.1 and fixed over repeated draws of $A$.

For estimator $j$ and domain $m$, relative bias (RB) and RRMSE were computed as
\begin{align}
 \operatorname{RB}_{jm}
 &=100\,\frac{R^{-1}\sum_{r=1}^{R}(\hat t_{jm}^{(r)}-t_m)}{t_m}, \label{eq:sim_rb_final}\\
 \operatorname{RRMSE}_{jm}
 &=100\,\frac{\left\{R^{-1}\sum_{r=1}^{R}(\hat t_{jm}^{(r)}-t_m)^2\right\}^{1/2}}{t_m}. \label{eq:sim_rrmse_final}
\end{align}
The reported results are based on $R=2\,000$ probability samples per scenario. Because the objective was to compare point estimators rather than variance estimators, the first-stage log-scale design variances supplied to the area-level models were estimated separately from an independent set of $1\,000$ probability samples. These domain-specific variances were treated as fixed in the $2\,000$-sample evaluation, avoiding a nested bootstrap. IPW proxy-total uncertainty was estimated using 500 replications of the pseudo-population bootstrap in Section~2.5.1.

\subsection{Simulation results}\label{sec:sim_results}

Table~\ref{tab:sim_results_final} reports, across the 60 domains, the median absolute relative bias and median RRMSE. Results for all 12 scenarios, signed biases, upper-tail summaries, MC feasibility rates and fallback reasons are provided in the supplementary material.

\begin{table}[H]
\centering
\small
\caption{Median absolute relative bias and median relative root mean squared error in the core scenarios (percent).}
\label{tab:sim_results_final}
\begin{tabular}{@{}c cc cc cc cc@{}}
\toprule
& \multicolumn{2}{c}{GREG}
& \multicolumn{2}{c}{MC}
& \multicolumn{2}{c}{EBLUP}
& \multicolumn{2}{c}{YL} \\
\cmidrule(lr){2-3}
\cmidrule(lr){4-5}
\cmidrule(lr){6-7}
\cmidrule(lr){8-9}
Scenario
& $|\mathrm{RB}|$ & RRMSE
& $|\mathrm{RB}|$ & RRMSE
& $|\mathrm{RB}|$ & RRMSE
& $|\mathrm{RB}|$ & RRMSE \\
\midrule
S1 & 0.144 & 1.933 & 0.054 & 2.183 & 0.349 & 2.149 & 0.529 & 1.801 \\
S2 & 0.144 & 1.933 & 0.177 & 2.260 & 0.403 & 2.194 & 0.529 & 1.801 \\
S3 & 0.144 & 1.933 & 0.067 & 2.010 & 0.311 & 1.935 & 0.492 & 1.815 \\
S4 & 0.144 & 1.933 & 0.179 & 2.083 & 0.367 & 1.981 & 0.492 & 1.815 \\
S5 & 0.144 & 1.933 & 0.034 & 1.621 & 0.214 & 1.547 & 0.561 & 1.799 \\
S6 & 0.144 & 1.933 & 0.093 & 2.391 & 0.457 & 2.280 & 0.396 & 1.834 \\
\bottomrule
\end{tabular}
\tabnote{Values are median absolute relative bias ($|\operatorname{RB}|$) and median relative root mean squared error (RRMSE) across domains. GREG = generalized regression estimator; MC = model-calibration estimator; EBLUP = empirical best linear unbiased predictor based on the operational first-stage estimator; YL = Ybarra--Lohr predictor (benchmark).}
\end{table}

The MC estimator remained nearly design-unbiased in all scenarios. Across the full set of 12 scenarios, its median absolute relative bias ranged from 0.022\% to 0.266\%, whereas its median RRMSE ranged from 1.442\% to 2.398\%. Misspecifying $g$ increased the median absolute bias and reduced efficiency, but did not generate substantial design bias. For example, under MAR and medium coverage, misspecification increased the median MC absolute bias from 0.054\% to 0.177\% and the median RRMSE from 2.183\% to 2.260\%. The corresponding EBLUPs displayed a more visible bias--variance trade-off due to area-level shrinkage.

Overlap was the main determinant of practical efficiency and stability. Under the correctly specified MAR scenarios, MC was feasible in 96.4\%, 90.6\% and 59.0\% of domain-replication combinations at high, medium and low coverage, respectively. At high coverage (S5), the median RRMSE fell from 1.933\% for GREG to 1.621\% for MC and 1.547\% for EBLUP, reductions of about 16\% and 20\%. At medium coverage (S1), the additional calibration variability outweighed the proxy-information gain, and at low coverage (S6) the median MC RRMSE increased to 2.391\%. In the low-overlap scenario, approximately 30\% of the domain-replication combinations used fallback because $A_m\cap B_m$ was too small, while a further 11\% failed the positivity requirement for the calibration weights.

The non-MAR mechanism affected the IPW auxiliary totals used by the YL benchmark. At medium coverage, their median absolute relative error increased from 1.99\% under MAR to 6.87\% under non-MAR, and the 90th percentile increased from 5.00\% to 15.99\%. The error also became directional: the median signed error was 0.29\% under MAR and 6.87\% under non-MAR. Its effect on the final YL predictor was more moderate, with median RRMSEs of 1.801\% and 1.815\%, because the area-level model used the direct estimates and fully observed area covariates together with the error-prone proxy total.

The comparison of the MAR and non-MAR MC results should not be interpreted as a causal comparison of the two selection mechanisms, because each mechanism produced a different fixed realization of $B$. The relevant robustness finding is that conditioning on the realized $B$, non-MAR selection did not induce appreciable design bias in the MC estimator. Overall, the simulation confirms that misspecification of the working function primarily affects efficiency rather than design bias, while the attainable precision gain depends critically on sufficient domain-level overlap and stable calibration equations.

\section{Conclusions}\label{sec:conclusions}

We considered a common official statistics setting in which a non-probability sample $B$ provides a proxy $y^\star$ that is informative about the study variable $y$ but is observed only for a subset of the population. To exploit such incomplete auxiliary information without modeling the (typically unknown) selection mechanism of the non-probability sample, we proposed a two-step framework that treats the realized sample as fixed auxiliary information and bases inference on the probability sample design. First, a unit-level measurement-error working model is estimated design-weightedly on the linked overlap $A\cap B$ and used to obtain predictions for units in $B$. These predictions are then incorporated through domain-specific model-calibration constraints, yielding approximately design-unbiased estimators of domain totals with design-based uncertainty assessment. Second, the resulting domain totals and their variance estimates are used as inputs to the area-level FH model with exactly known covariates, producing EBLUPs that borrow strength across domains. The approach is flexible in the choice of the working model linking $y$ and $y^\star$ and can accommodate linear, nonlinear, or nonparametric relationships.

Empirical results from the three applications indicate that, within areas where model calibration is feasible, the MC step accounts for most of the improvement in domain-level precision, while the FH model provides an additional, generally smaller gain. The all-area operational assessment confirms that the improvements persist after including fallback areas. In that assessment, the FH contribution is especially visible in the upper tail for job vacancies, where direct first-stage estimates and their variances are used for municipalities without MC estimates. Large improvements are observed for both turnover and job vacancies, whereas the gains for investment are more moderate. Their ordering is not monotone in the relative size of the non-probability sample, indicating that the amount of proxy information, its unit-level predictive relationship with $y$, and the baseline precision of the direct estimator all matter. The Ybarra--Lohr benchmark exhibits more tail-focused and heterogeneous gains: its median RRMSE remains close to the direct estimator, although it reduces upper-tail unreliability. Overall, incorporating the proxy through unit-level prediction and MC constraints provides the most consistent precision improvement across the applications, with the FH step acting as a complementary smoothing stage and extending the procedure to areas where direct estimates replace unavailable MC inputs.

The design-based simulation supports these interpretations. Across 12 scenarios, the MC estimator remained approximately design-unbiased when the working function was misspecified and when selection into $B$ was non-MAR; misspecification primarily reduced efficiency rather than inducing substantial bias. Precision gains depended strongly on domain-level overlap. Under high coverage, MC and EBLUP reduced median RRMSE relative to GREG by about 16\% and 20\%, respectively, whereas low overlap increased fallback rates and RRMSE. Non-MAR selection substantially increased the error of the IPW proxy totals used by the YL benchmark, although area-level shrinkage moderated its effect on the final predictor. 

The proposed framework requires sufficiently large overlaps $A\cap B$ within domains; when the overlap is small, the design-weighted working model and the calibration constraints may become unstable or infeasible, and the MC estimator may be unavailable for some domains. The achievable gains also depend on the quality of the proxy and on the predictive adequacy of the working model.

\section*{Acknowledgements}

This project has received funding from the Research Council of Lithuania (LMTLT), agreement No~S-MIP-25-10.

\section*{Supplementary material}

The supplementary material reports complete results for all 12 simulation scenarios, including signed and absolute relative bias, upper-tail RRMSE summaries, model-calibration feasibility and fallback diagnostics, and errors of the IPW proxy totals.

\bibliographystyle{apalike}      
\bibliography{literature}

\clearpage
\pdfbookmark[0]{Supplementary material}{supplement}
\section*{Supplementary material}

\counterwithout{equation}{section}
\counterwithout{table}{section}
\counterwithout{figure}{section}
\setcounter{section}{0}
\setcounter{subsection}{0}
\setcounter{equation}{0}
\setcounter{table}{0}
\setcounter{figure}{0}
\renewcommand{\thesection}{S.\arabic{section}}
\renewcommand{\thesubsection}{S.\arabic{section}.\arabic{subsection}}
\renewcommand{\thesubsubsection}{\thesubsection.\arabic{subsubsection}}
\renewcommand{\thetable}{S.\arabic{table}}
\renewcommand{\thefigure}{S.\arabic{figure}}
\renewcommand{\theequation}{S.\arabic{equation}}
\renewcommand{\theHsection}{supp.\arabic{section}}
\renewcommand{\theHsubsection}{\theHsection.\arabic{subsection}}
\renewcommand{\theHsubsubsection}{\theHsubsection.\arabic{subsubsection}}
\renewcommand{\theHtable}{supp.\arabic{table}}
\renewcommand{\theHfigure}{supp.\arabic{figure}}
\renewcommand{\theHequation}{supp.\arabic{equation}}
\section{Purpose and scope}

This supplement provides the full design and results of the focused design-based simulation reported in Section~4 of the main paper. The main paper presents six core scenarios. Here we report the complete $2\times2\times3$ factorial experiment, additional bias and upper-tail summaries, model-calibration feasibility and fallback diagnostics, and diagnostics for the inverse-probability-weighted proxy totals used by the Ybarra--Lohr benchmark.

A single finite population was held fixed throughout the experiment. For each combination of selection mechanism and target coverage, one realization of the non-probability sample $B$ was generated, held fixed, and shared by the two working-function specifications. Repeated probability samples $A$ were then selected under the known probability design. The reported Monte Carlo properties are therefore conditional on the fixed finite population and the corresponding realized $B$ and refer to repeated sampling with respect to $A$.

\section{Additional simulation details}

\subsection{Finite population and probability sampling design}

The finite population contained $M=60$ domains and $21\,600$ units. Domain sizes were rounded values on an equally spaced sequence from 220 to 500. Independently generated domain effects $a_m,u_m\sim\mathcal{N}(0,1)$ and unit-level errors $\xi_{mi},v_{mi}\sim\mathcal{N}(0,1)$ and $\varepsilon_{mi}\sim\mathcal{N}(0,4^2)$ were used to generate
\begin{align}
x_{mi} &= a_m+\xi_{mi},\\
y_{mi}^{\star} &= 5+1.2x_{mi}+0.6a_m+v_{mi},\\
y_{mi} &= 50+3x_{mi}+7y_{mi}^{\star}
          +1.5(y_{mi}^{\star}-5)^2+4u_m+\varepsilon_{mi}.
\end{align}
The 60 domains coincided with the sampling strata. A simple random sample without replacement of $n_m=30$ units was drawn independently within every domain/stratum. Thus the total probability-sample size was $1\,800$ and the design weight in domain $m$ was $d_{mi}=N_m/n_m$.

The direct estimator was a domain generalized regression (GREG) estimator using the known domain size and the known domain total of $x$. The fully observed area-level covariates in the Fay--Herriot (FH) models were $\bz_m=(1,\log N_m,\bar x_m)^\T$, where $\bar x_m=N_m^{-1}\sum_{i\in U_m}x_{mi}$.

\subsection{Non-probability sample and working functions}

The non-probability sample indicator was generated from
\begin{equation}
\PP(\delta_{mi}=1)
=
\operatorname{logit}^{-1}\!\left\{
\alpha_{\rho}+0.7x_{mi}-0.4\bar x_m+\lambda r_{mi}
\right\},
\qquad
r_{mi}=\frac{v_{mi}+\varepsilon_{mi}/4}{\sqrt{2}},
\end{equation}
where $\alpha_{\rho}$ was chosen to attain target coverage
$\rho\in\{0.20,0.40,0.70\}$. Under the missing-at-random (MAR) mechanism, $\lambda=0$. Under the non-MAR mechanism, $\lambda=1$, while the fitted propensity score model continued to use only the fully observed frame variables $x_{mi}$ and $\bar x_m$.

The correctly specified working function was
\begin{equation}
g_{\mathrm C}(y^{\star},x)
=
\theta_0+\theta_1x+\theta_2y^{\star}
+\theta_3(y^{\star}-5)^2,
\end{equation}
and the misspecified version omitted the quadratic term,
\begin{equation}
g_{\mathrm M}(y^{\star},x)
=
\theta_0+\theta_1x+\theta_2y^{\star}.
\end{equation}
Both functions were re-estimated from $A\cap B$ by design-weighted least squares in every probability-sample replication.

\subsection{Operational estimation and feasibility rules}

Four estimators were evaluated: the direct GREG estimator, the operational model-calibration (MC) estimator, the corresponding empirical best linear unbiased predictor (EBLUP), and the Ybarra--Lohr (YL) predictor. If model calibration was feasible, the first-stage estimate for the Fay--Herriot model was the model-calibration estimate. Otherwise, the direct estimate and its design-based variance were used.

In the simulation, model calibration was used only when all of the following numerical conditions held:
\begin{itemize}[leftmargin=1.8em,itemsep=0.15em,topsep=0.25em]
\item at least five sampled units belonged to $A_m\cap B_m$;
\item at least two sampled units belonged to $A_m\setminus B_m$;
\item at least two distinct fitted values were available in $A_m\cap B_m$;
\item the cross-product matrix formed from the three model-calibration variables---the two post-stratum indicators and the fitted proxy value---was of full rank and sufficiently far from singular for stable numerical inversion; operationally, its condition number did not exceed $10^{10}$;
\item all calibration weights were positive and the absolute ratio of each calibrated weight to its design weight, $|w_i/d_i|$, did not exceed 25.
\end{itemize}
When a condition failed, the direct estimator was used as the operational first-stage estimator in that domain.

For the YL predictor, the propensity score model was fitted from the full finite-population frame. The domain proxy total and its log-scale measurement-error variance were obtained using, respectively, the normalized inverse probability weighted (IPW) estimator and the pseudo-population bootstrap described in Section~2.5.1 of the main paper; the bootstrap used 500 replications.

\subsection{Monte Carlo implementation}

For each domain, the direct-estimator log-scale design variance and, separately for each scenario, the model-calibration log-scale design variance conditional on successful calibration were estimated as empirical variances over an independent set of $1\,000$ probability samples. During the reported evaluation, the corresponding fixed variance was used according to whether model calibration was feasible in a given sample. Thus these variances were not re-estimated from each of the $2\,000$ evaluation samples. This deliberate known-variance approximation isolates point-estimator performance and avoids a nested bootstrap; it does not assess the additional variability caused by sample-specific variance estimation.

The reported results are based on $2\,000$ probability samples per scenario. Table~\ref{tab:supp_scenarios} summarizes all 12 scenarios and the corresponding design diagnostics.

\begin{table}[!ht]
\centering
\footnotesize
\caption{Full set of simulation scenarios and design diagnostics.}
\label{tab:supp_scenarios}
\begin{tabular}{@{}l l l c c c c c c@{}}
\toprule
Scenario & Working & Selection & Target & Realized & Median & \multicolumn{3}{c}{MC feasibility (\%)}\\
\cmidrule(lr){7-9}
 & function & into $B$ & coverage & coverage & overlap & Mean & P10 & Median\\
\midrule
S1  & Correct       & MAR     & 0.40 & 0.395 & 11.5 & 90.6 & 84.1 & 91.9 \\
S2  & Misspecified  & MAR     & 0.40 & 0.395 & 11.5 & 90.9 & 87.6 & 91.7 \\
S3  & Correct       & Non-MAR & 0.40 & 0.400 & 11.8 & 91.0 & 86.8 & 92.0 \\
S4  & Misspecified  & Non-MAR & 0.40 & 0.400 & 11.8 & 91.3 & 88.2 & 92.0 \\
S5  & Correct       & MAR     & 0.70 & 0.700 & 20.7 & 96.4 & 93.8 & 96.6 \\
S6  & Correct       & MAR     & 0.20 & 0.199 &  5.9 & 59.0 & 37.7 & 63.4 \\
S7  & Correct       & Non-MAR & 0.70 & 0.698 & 21.1 & 96.3 & 94.6 & 96.5 \\
S8  & Correct       & Non-MAR & 0.20 & 0.197 &  6.0 & 59.1 & 33.1 & 64.7 \\
S9  & Misspecified  & MAR     & 0.70 & 0.700 & 20.7 & 96.4 & 95.1 & 96.3 \\
S10 & Misspecified  & MAR     & 0.20 & 0.199 &  5.9 & 59.6 & 38.0 & 64.1 \\
S11 & Misspecified  & Non-MAR & 0.70 & 0.698 & 21.1 & 96.4 & 95.5 & 96.5 \\
S12 & Misspecified  & Non-MAR & 0.20 & 0.197 &  6.0 & 59.3 & 34.1 & 64.2 \\
\bottomrule
\end{tabular}
\tabnote{S1--S6 are the core scenarios reported in the main paper; S7--S12 complete the factorial design. Median overlap is the median expected value of $|A_m\cap B_m|$ across domains. P10 (the first decile) and Median refer to the distribution across domains of the domain-specific model-calibration feasibility rate. MC = model-calibration estimator; MAR = missing-at-random.}
\end{table}

\section{Full performance results}

For estimator $j$ and domain $m$, relative bias (RB) and relative root mean squared error (RRMSE) were
\begin{align}
\operatorname{RB}_{jm}
&=
100\,\frac{R^{-1}\sum_{r=1}^{R}
(\hat t_{jm}^{(r)}-t_m)}{t_m},\\
\operatorname{RRMSE}_{jm}
&=
100\,\frac{
\left\{R^{-1}\sum_{r=1}^{R}
(\hat t_{jm}^{(r)}-t_m)^2\right\}^{1/2}
}{t_m}.
\end{align}
Tables~\ref{tab:supp_bias} and~\ref{tab:supp_rrmse} provide the complete results across all 12 scenarios and all four estimators.

\begingroup
\fontsize{10}{11}\selectfont
\renewcommand{\arraystretch}{0.90}
\setlength{\tabcolsep}{7pt}
\setlength{\LTpre}{12pt}
\setlength{\LTpost}{0pt}
\begin{longtable}{@{}l l r r r@{}}
\caption{Bias summaries across the 60 domains (percent).}
\label{tab:supp_bias}\\
\toprule
Scenario & Estimator & Median RB & Median $|\mathrm{RB}|$ & P90 $|\mathrm{RB}|$\\
\midrule
\endfirsthead
\multicolumn{5}{l}{\textit{Table \thetable\ continued}}\\
\toprule
Scenario & Estimator & Median RB & Median $|\mathrm{RB}|$ & P90 $|\mathrm{RB}|$\\
\midrule
\endhead
\midrule
\multicolumn{5}{r}{\textit{Continued on next page}}\\
\endfoot
\bottomrule
\endlastfoot
S1 & Direct & -0.144 & 0.144 & 0.207 \\
S1 & MC & -0.051 & 0.054 & 0.123 \\
S1 & EBLUP & -0.049 & 0.349 & 0.977 \\
S1 & YL & -0.062 & 0.529 & 1.034 \\
S2 & Direct & -0.144 & 0.144 & 0.207 \\
S2 & MC & -0.177 & 0.177 & 0.244 \\
S2 & EBLUP & -0.193 & 0.403 & 0.962 \\
S2 & YL & -0.062 & 0.529 & 1.034 \\
S3 & Direct & -0.144 & 0.144 & 0.207 \\
S3 & MC & -0.067 & 0.067 & 0.116 \\
S3 & EBLUP & -0.058 & 0.311 & 0.929 \\
S3 & YL & 0.010 & 0.492 & 1.016 \\
S4 & Direct & -0.144 & 0.144 & 0.207 \\
S4 & MC & -0.179 & 0.179 & 0.239 \\
S4 & EBLUP & -0.189 & 0.367 & 0.930 \\
S4 & YL & 0.010 & 0.492 & 1.016 \\
S5 & Direct & -0.144 & 0.144 & 0.207 \\
S5 & MC & -0.006 & 0.034 & 0.069 \\
S5 & EBLUP & -0.013 & 0.214 & 0.595 \\
S5 & YL & -0.064 & 0.561 & 0.976 \\
S6 & Direct & -0.144 & 0.144 & 0.207 \\
S6 & MC & -0.093 & 0.093 & 0.211 \\
S6 & EBLUP & -0.105 & 0.457 & 1.185 \\
S6 & YL & -0.110 & 0.396 & 0.945 \\
S7 & Direct & -0.144 & 0.144 & 0.207 \\
S7 & MC & -0.013 & 0.022 & 0.070 \\
S7 & EBLUP & -0.015 & 0.187 & 0.472 \\
S7 & YL & -0.116 & 0.579 & 1.039 \\
S8 & Direct & -0.144 & 0.144 & 0.207 \\
S8 & MC & -0.237 & 0.237 & 0.309 \\
S8 & EBLUP & -0.211 & 0.481 & 0.914 \\
S8 & YL & -0.074 & 0.510 & 1.005 \\
S9 & Direct & -0.144 & 0.144 & 0.207 \\
S9 & MC & -0.193 & 0.193 & 0.269 \\
S9 & EBLUP & -0.177 & 0.351 & 0.673 \\
S9 & YL & -0.064 & 0.561 & 0.976 \\
S10 & Direct & -0.144 & 0.144 & 0.207 \\
S10 & MC & -0.143 & 0.143 & 0.231 \\
S10 & EBLUP & -0.140 & 0.449 & 1.126 \\
S10 & YL & -0.110 & 0.396 & 0.945 \\
S11 & Direct & -0.144 & 0.144 & 0.207 \\
S11 & MC & -0.198 & 0.198 & 0.258 \\
S11 & EBLUP & -0.165 & 0.286 & 0.622 \\
S11 & YL & -0.116 & 0.579 & 1.039 \\
S12 & Direct & -0.144 & 0.144 & 0.207 \\
S12 & MC & -0.266 & 0.266 & 0.345 \\
S12 & EBLUP & -0.242 & 0.495 & 0.928 \\
S12 & YL & -0.074 & 0.510 & 1.005 \\
\end{longtable}
\tabnote{RB = relative bias; P90 = the ninth decile; MC = model-calibration estimator; EBLUP = empirical best linear unbiased predictor; YL = Ybarra--Lohr predictor.}
\endgroup
\clearpage

\begingroup
\fontsize{10}{11}\selectfont
\renewcommand{\arraystretch}{0.90}
\setlength{\tabcolsep}{7pt}
\setlength{\LTpre}{12pt}
\setlength{\LTpost}{0pt}
\begin{longtable}{@{}l l r r r r@{}}
\caption{Relative root mean squared error summaries across the 60 domains (percent).}
\label{tab:supp_rrmse}\\
\toprule
Scenario & Estimator & Median & P75 & P90 & Maximum\\
\midrule
\endfirsthead
\multicolumn{6}{l}{\textit{Table \thetable\ continued}}\\
\toprule
Scenario & Estimator & Median & P75 & P90 & Maximum\\
\midrule
\endhead
\midrule
\multicolumn{6}{r}{\textit{Continued on next page}}\\
\endfoot
\bottomrule
\endlastfoot
S1 & Direct & 1.933 & 2.231 & 2.377 & 2.689 \\
S1 & MC & 2.183 & 2.449 & 2.612 & 2.871 \\
S1 & EBLUP & 2.149 & 2.253 & 2.454 & 3.097 \\
S1 & YL & 1.801 & 1.971 & 2.165 & 2.911 \\
S2 & Direct & 1.933 & 2.231 & 2.377 & 2.689 \\
S2 & MC & 2.260 & 2.524 & 2.652 & 2.906 \\
S2 & EBLUP & 2.194 & 2.302 & 2.478 & 3.177 \\
S2 & YL & 1.801 & 1.971 & 2.165 & 2.911 \\
S3 & Direct & 1.933 & 2.231 & 2.377 & 2.689 \\
S3 & MC & 2.010 & 2.260 & 2.420 & 2.622 \\
S3 & EBLUP & 1.935 & 2.132 & 2.217 & 2.498 \\
S3 & YL & 1.815 & 1.984 & 2.176 & 2.663 \\
S4 & Direct & 1.933 & 2.231 & 2.377 & 2.689 \\
S4 & MC & 2.083 & 2.290 & 2.454 & 2.656 \\
S4 & EBLUP & 1.981 & 2.162 & 2.298 & 2.513 \\
S4 & YL & 1.815 & 1.984 & 2.176 & 2.663 \\
S5 & Direct & 1.933 & 2.231 & 2.377 & 2.689 \\
S5 & MC & 1.621 & 1.782 & 1.858 & 2.086 \\
S5 & EBLUP & 1.547 & 1.726 & 1.828 & 2.019 \\
S5 & YL & 1.799 & 2.007 & 2.128 & 2.703 \\
S6 & Direct & 1.933 & 2.231 & 2.377 & 2.689 \\
S6 & MC & 2.391 & 2.786 & 3.018 & 3.292 \\
S6 & EBLUP & 2.280 & 2.469 & 2.754 & 3.559 \\
S6 & YL & 1.834 & 2.015 & 2.167 & 2.444 \\
S7 & Direct & 1.933 & 2.231 & 2.377 & 2.689 \\
S7 & MC & 1.442 & 1.585 & 1.731 & 1.936 \\
S7 & EBLUP & 1.384 & 1.534 & 1.717 & 1.957 \\
S7 & YL & 1.833 & 2.003 & 2.126 & 2.621 \\
S8 & Direct & 1.933 & 2.231 & 2.377 & 2.689 \\
S8 & MC & 2.307 & 2.697 & 2.838 & 3.050 \\
S8 & EBLUP & 2.226 & 2.385 & 2.557 & 3.221 \\
S8 & YL & 1.849 & 1.978 & 2.194 & 2.637 \\
S9 & Direct & 1.933 & 2.231 & 2.377 & 2.689 \\
S9 & MC & 1.775 & 1.944 & 2.055 & 2.198 \\
S9 & EBLUP & 1.766 & 1.852 & 1.949 & 2.335 \\
S9 & YL & 1.799 & 2.007 & 2.128 & 2.703 \\
S10 & Direct & 1.933 & 2.231 & 2.377 & 2.689 \\
S10 & MC & 2.398 & 2.809 & 3.021 & 3.328 \\
S10 & EBLUP & 2.289 & 2.474 & 2.779 & 3.608 \\
S10 & YL & 1.834 & 2.015 & 2.167 & 2.444 \\
S11 & Direct & 1.933 & 2.231 & 2.377 & 2.689 \\
S11 & MC & 1.630 & 1.737 & 1.907 & 2.053 \\
S11 & EBLUP & 1.595 & 1.676 & 1.833 & 2.408 \\
S11 & YL & 1.833 & 2.003 & 2.126 & 2.621 \\
S12 & Direct & 1.933 & 2.231 & 2.377 & 2.689 \\
S12 & MC & 2.315 & 2.707 & 2.830 & 3.090 \\
S12 & EBLUP & 2.233 & 2.400 & 2.536 & 3.285 \\
S12 & YL & 1.849 & 1.978 & 2.194 & 2.637 \\
\end{longtable}
\tabnote{P75 = the third quartile; P90 = the ninth decile; MC = model-calibration estimator; EBLUP = empirical best linear unbiased predictor; YL = Ybarra--Lohr predictor.}
\endgroup
\clearpage

Across all 12 scenarios, the median absolute relative bias of the model-calibration estimator ranged from 0.022\% to 0.266\%, while its median RRMSE ranged from 1.442\% to 2.398\%. Misspecification of the working function generally increased the absolute bias and RRMSE, particularly at high coverage, but the model-calibration bias remained small relative to total error.

At high coverage, the model-calibration estimator improved on GREG under both MAR and non-MAR realizations. At low coverage, the main limitation was not design bias but insufficient overlap and unstable calibration. The EBLUP exhibited a more visible bias--variance trade-off because of area-level shrinkage.

\section{Model-calibration feasibility and fallback}

Table~\ref{tab:supp_fallback} separates successful model calibration from the principal fallback reasons. The proportions are calculated over all $60\times 2\,000$ domain-replication combinations in each scenario.

\begin{table}[!ht]
\centering
\small
\caption{Model-calibration use and fallback reasons (percent of domain-replication combinations).}
\label{tab:supp_fallback}
\begin{tabular}{@{}l r r r r@{}}
\toprule
Scenario & MC used & Insufficient $A_m\cap B_m$ & Insufficient $A_m\setminus B_m$ & Nonpositive weights\\
\midrule
S1 & 90.6 & 0.7 & 0.0 & 8.7 \\
S2 & 90.9 & 0.7 & 0.0 & 8.4 \\
S3 & 91.0 & 0.4 & 0.0 & 8.6 \\
S4 & 91.3 & 0.4 & 0.0 & 8.2 \\
S5 & 96.4 & 0.0 & 0.2 & 3.4 \\
S6 & 59.0 & 30.1 & 0.0 & 10.9 \\
S7 & 96.3 & 0.0 & 0.1 & 3.6 \\
S8 & 59.1 & 30.6 & 0.0 & 10.3 \\
S9 & 96.4 & 0.0 & 0.2 & 3.4 \\
S10 & 59.6 & 30.1 & 0.0 & 10.3 \\
S11 & 96.4 & 0.0 & 0.1 & 3.5 \\
S12 & 59.3 & 30.6 & 0.0 & 10.1 \\
\bottomrule
\end{tabular}
\tabnote{Other numerical failure reasons did not occur in the experiment. Percentages may differ from 100 by rounding. MC = model-calibration estimator.}
\end{table}

Under low coverage, insufficient overlap accounted for approximately 30\% of domain-replication combinations and nonpositive calibration weights for a further 10--11\%. Under medium coverage, insufficient overlap was rare and nonpositive weights were the dominant fallback reason. Under high coverage, model calibration was feasible in approximately 96\% of combinations.

\section{Diagnostics for proxy totals estimated by inverse probability weighting}

The realized non-probability sample was the same for the correct and misspecified working-function scenarios with a given selection mechanism and coverage. Table~\ref{tab:supp_ipw} therefore reports six distinct sets of diagnostics. The relative error is calculated against the true finite-population proxy total in each domain.

\begin{table}[!ht]
\centering
\small
\caption{Diagnostics for domain proxy totals estimated by inverse probability weighting.}
\label{tab:supp_ipw}
\setlength{\tabcolsep}{5.5pt}
\begin{tabular}{@{}l l r r r r r@{}}
\toprule
Selection & Coverage & Realized & Median signed & Median absolute & P90 absolute & Median $\hat C_{m\star}$\\
mechanism & level & coverage & error (\%) & error (\%) & error (\%) & (log scale)\\
\midrule
MAR & High & 0.700 & 0.079 & 0.806 & 2.857 & 0.000143 \\
MAR & Medium & 0.395 & 0.290 & 1.993 & 4.999 & 0.000521 \\
MAR & Low & 0.199 & 0.636 & 2.930 & 12.024 & 0.001457 \\
non-MAR & High & 0.698 & 3.463 & 3.463 & 7.691 & 0.000110 \\
non-MAR & Medium & 0.400 & 6.869 & 6.869 & 15.992 & 0.000349 \\
non-MAR & Low & 0.197 & 10.590 & 10.590 & 22.582 & 0.001017 \\
\bottomrule
\end{tabular}
\tabnote{$\hat C_{m\star}$ is the estimated log-scale measurement-error variance associated with the inverse-probability-weighted proxy total, using the notation of Section~2.5 of the main paper. P90 = the ninth decile; MAR = missing-at-random.}
\end{table}

The non-MAR mechanism produced a clearly directional error in the IPW proxy totals. At medium coverage, the median absolute error increased from 1.993\% under MAR to 6.869\% under non-MAR, and the P90 increased from 4.999\% to 15.992\%. The effect on the final Ybarra--Lohr predictor was more moderate because the area-level model also used the direct estimator and fully observed domain covariates.

\section{Interpretation of the full factorial experiment}

The additional six scenarios reinforce the conclusions drawn from the core scenarios.

\begin{itemize}[leftmargin=1.8em]
\item \textbf{Working-function misspecification.} At high coverage, misspecification increased the median model-calibration RRMSE from 1.621\% to 1.775\% under MAR and from 1.442\% to 1.630\% under non-MAR. The corresponding median absolute biases remained below 0.20\%.
\item \textbf{Overlap.} High coverage allowed the proxy information to reduce sampling variability. At low coverage, frequent fallback and unstable calibration weights dominated the results, so the operational estimator was less efficient even though its design bias remained small.
\item \textbf{Non-MAR selection.} The non-MAR mechanism substantially affected the IPW proxy totals. Its effect on the final Ybarra--Lohr predictor was moderated by area-level shrinkage. Comparisons of model-calibration RRMSE between MAR and non-MAR scenarios should not be interpreted causally because each mechanism generated a different fixed realization of $B$.
\item \textbf{Bias--variance trade-off.} For model calibration, differences in RRMSE were driven primarily by sampling variability rather than design bias. EBLUP and Ybarra--Lohr predictions showed a more visible bias--variance trade-off because they use area-level shrinkage.
\end{itemize}

\section{Data and code availability}

The simulation used only synthetic data generated according to Section~S.2.1; no external or confidential data were required. The supplementary material accompanying the paper consists of this document, and no separate code archive is included.

\end{document}